\documentclass[prd,floatfix,showpacs]{revtex4}
\usepackage{epsfig}

\newcommand{\eq}[1]{(\ref{#1})}

\newcommand{\be}{\begin{equation}}
\newcommand{\ee}{\end{equation}}
\newcommand{\bea}{\begin{eqnarray}}
\newcommand{\eea}{\end{eqnarray}}
\newcommand{\ben}{\begin{eqnarray*}}
\newcommand{\een}{\end{eqnarray*}}

\newcommand{\DS}{Dyson--Schwinger }

\newcommand{\w}{\omega}
\newcommand{\e}{\varepsilon}
\newcommand{\al}{\alpha}
\newcommand{\ba}{\beta}
\newcommand{\ga}{\gamma}
\newcommand{\G}{\Gamma}
\newcommand{\de}{\delta}

\newcommand{\si}{\sigma}

\newcommand{\ro}{\rho}
\newcommand{\la}{\lambda}

\newcommand{\ta}{\tau}

\newcommand{\ha}{\frac{1}{2}}
\newcommand{\pd}{\partial}

\newcommand{\cd}{{\cal D}}
\newcommand{\cs}{{\cal S}}
\renewcommand{\div}{\vec{\nabla}}

\newcommand{\s}[2]{{#1}\!\cdot\!{#2}}
\newcommand{\ov}[1]{\overline{#1}}
\newcommand{\dk}[1]{\,\,\,\raisebox{-0.4ex}{\large $\bar{}$}\!\!d\,{#1}\,}

\begin{document}
\title{Perturbation Theory of Coulomb Gauge Yang-Mills Theory Within the 
First Order Formalism}
\author{P.~Watson}
\author{H.~Reinhardt}
\affiliation{Institut f\"ur Theoretische Physik, Universit\"at T\"ubingen,
 Auf der Morgenstelle 14, D-72076 T\"ubingen, Deutschland}
\begin{abstract}
Perturbative Coulomb gauge Yang-Mills theory within the first order 
formalism is considered.  Using a differential equation technique 
and dimensional regularization, analytic results for both the ultraviolet 
divergent and finite parts of the two-point functions at one-loop order are 
derived.  It is shown how the non-ultraviolet divergent parts of the results 
are finite at spacelike momenta with kinematical singularities on the 
light-cone and subsequent branch cuts extending into the timelike region.
\end{abstract}
\maketitle
\section{Introduction}
\setcounter{equation}{0}
Coulomb gauge QCD is rather special.  Amongst all the various gauges, it can 
be shown in Coulomb gauge that the number of dynamical variables reduces to 
the number of physical degrees of freedom \cite{Zwanziger:1998ez}.  This 
allows for a tantalizing glimpse of possible nonperturbative descriptions 
of confinement and the hadron spectrum.  The so-called ``Gribov--Zwanziger" 
scenario of confinement, \cite{Gribov:1977wm,Zwanziger:1998ez}, becomes 
especially relevant in Coulomb gauge.  In this picture, the temporal 
component of the gluon propagator provides for a long-range confining 
force whilst the transverse spatial components are suppressed in the 
infrared.  Various calculations support this picture, among them 
\cite{Szczepaniak:2001rg,Greensite:2003xf,Zwanziger:2003de,Feuchter:2004mk,
Nakamura:2005ux}.

Perhaps a touch ironically, perturbation theory is one of the starting 
ingredients for nonperturbative calculations in the sense that in the 
high energy region (and for asymptotically free theories such as QCD 
this is the perturbative domain), the regularization and renormalization 
of the theory play an important role in constraining the necessary 
approximations.  In Coulomb gauge, only the leading divergence structure 
is known due to severe technical difficulties (see for example 
\cite{Heinrich:1999ak,Andrasi:2005xu}).  For Coulomb gauge within the 
(standard) second order formalism there exist so-called energy divergences, 
which have been shown to cancel up to two-loops \cite{Andrasi:2005xu}, but 
a general proof of this cancellation is sadly lacking.  One way to 
circumvent the energy divergences is to work within the first order 
formalism, where formal arguments show that such divergences cancel exactly 
\cite{Zwanziger:1998ez}.  This circumvention comes at a price: the \DS 
equations become cumbersome \cite{Watson:2006yq} and full multiplicative 
renormalizability is not maintained \cite{Zwanziger:1998ez}.  Whilst the 
leading divergences do provide crucial information about the renormalization 
of the theory, the remaining finite parts are critical to further progress 
in the field.  Having expounded the necessity for perturbative results 
within the nonperturbative context, obviously perturbative results for 
physical high energy processes are desirable -- not the least in order to 
compare with results from covariant gauges.

The technical barrier to progress in Coulomb gauge perturbation theory stems 
from noncovariant loop integrals of the type:
\be
\int\frac{d^4\w}{\w^2(k-\w)^2\vec{\w}^2(\vec{k}-\vec{\w})^2}
\ee
where (in Euclidean space) $\w^2=\w_4^2+\vec{\w}^2$.  Standard techniques 
such as Schwinger parametrization \cite{collins} fail due to the complexity 
of the resulting parametric integrals.  One might imagine that using contour 
integration to firstly evaluate the temporal component of the integral might 
make the situation simpler.  However, in such a method, translational 
invariance is lost and since the subsequent spatial integral is ultraviolet 
[UV] divergent, the result will in general be incorrect.  A UV-cutoff 
procedure will also fail.  There are however techniques that can overcome 
these difficulties and one of these is the differential equation technique.  
In its original form \cite{Kotikov:1990kg,Argeri:2007up}, complicated 
massive integrals arising in covariant gauge calculations can be considered 
and when supplemented with integration by parts identities 
\cite{Chetyrkin:1981qh,Tkachov:1981wb}, the technique becomes a powerful 
tool.  The ethos of the technique is that whilst the multi--dimensional 
parametric form of the integral may be practically impossible to work with, 
the original integral is itself only a function of a few variables and where 
differential equations can be derived, finding the solution involves 
integration over only these few variables and sorting out the boundary 
conditions.

In this paper, we consider the one-loop perturbative two-point functions of 
Coulomb gauge Yang-Mills theory within the first order formalism.  Within 
this noncovariant setting, a variant of the differential equation technique 
and the integration by parts identities are derived in order to evaluate 
integrals such as the one above.  This allows for a full analysis of the 
various propagator and two-point proper functions of the theory.

The paper is organized as follows.  We start by briefly reviewing the first 
order formalism used and express the two-point functions in terms of their 
loop integrals.  In Section~3, the noncovariant integrals inherent to 
Coulomb gauge are evaluated.  This section comprises the bulk of the 
development necessary to the study and is unashamedly technical in nature.  
The results for the two-point functions are collected in Section~4.  We 
finish with a summary and outlook.  Those loop integrals that can be 
evaluated using standard techniques are described in 
Appendix~\ref{app:int0}.  Appendix~\ref{app:intcheck} contains a nontrivial 
check on the noncovariant integrals.

\section{The First order formalism and Perturbation Theory}
\setcounter{equation}{0}
Since Coulomb gauge Yang-Mills theory within the first order formalism is 
rather different to Yang-Mills theory in linear covariant gauges, let us 
begin by reviewing those aspects of the formalism that will be relevant.  
For a complete description, the reader is referred to 
Ref.~\cite{Watson:2006yq}.  The generating functional is written 
(in Minkowski space)
\be
Z[J]=\int\cd\Phi\exp{\left\{\imath\cs_B+\imath\cs_{fp}+\imath\cs_{\pi}
+\imath\cs_s\right\}}
\label{eq:gen0}
\ee
where $\Phi$ denotes the collection of fields and the terms in the action 
are given by
\bea
\cs_B&=&\int d^4x\left[-\ha\s{\vec{B}^a}{\vec{B}^a}\right],\nonumber\\
\cs_{fp}&=&\int d^4x\left[-\la^a\s{\vec{\nabla}}{\vec{A}^a}
-\ov{c}^a\s{\vec{\nabla}}{\vec{D}^{ab}}c^b\right],\nonumber\\
\cs_{\pi}&=&\int d^4x\left[-\ta^a\s{\vec{\nabla}}{\vec{\pi}^a}
-\ha\s{(\vec{\pi}^a-\div\phi^a)}{(\vec{\pi}^a-\div\phi^a)}
+\s{(\vec{\pi}^a-\div\phi^a)}{\left(\pd^0\vec{A}^a+\vec{D}^{ab}\si^b
\right)}\right]
\label{eq:act}
\eea
with the source term defined in condensed notation as (Greek indices such 
as $\al$ refer to all attributes of the field, including its type, and 
summation over all discrete indices and integration over all continuous 
arguments is implicitly understood):
\be
\cs_s=J_\al\Phi_\al.
\label{eq:source0}
\ee
In the above, $\vec{A}$ and $\si$ are the spatial and temporal components 
of the gauge field, $\vec{\pi}$ and $\phi$ arise in the construction of 
the first order formalism (they represent the transverse and longitudinal 
components of the conjugate momentum to the gauge field), $\ov{c}$ and $c$ 
are the Grassmann-valued Faddeev--Popov ghost fields introduced by fixing 
the gauge, $\la$ and $\ta$ are Lagrange multiplier fields.  The 
chromomagnetic field, $\vec{B}$, is given by
\be
B_i^a=\epsilon_{ijk}\left[\nabla_jA_k^a-\ha gf^{abc}A_j^bA_k^c\right]
\ee
(roman subscripts indicate spatial indices) and the spatial component of 
the covariant derivative in the adjoint representation is
\be
\vec{D}^{ab}=\de^{ab}\div-gf^{acb}\vec{A}^c.
\ee

The general forms of the various Green's functions that we will be 
considering are constrained in several ways.  Since the derivation is 
necessarily somewhat longwinded, for brevity we omit the details here and 
again refer the reader to Ref.~\cite{Watson:2006yq} for a full account.  
There are three constraints.  Firstly, the Lagrange multiplier 
($\la$, $\ta$) and $\phi$ field equations of motion can be solved exactly.  
The former primarily supply the transversality properties of the 
vector--vector propagators (i.e., the connected two-point Green's 
functions), the latter relating the proper two-point functions involving 
functional derivatives of the $\phi$-field to contractions of those 
involving the corresponding $\vec{\pi}$-field.  Secondly, the equation 
stemming from the invariance of the generating functional under the BRS 
transform (the Ward--Takahashi identity in raw functional form) tells us 
that the $\la$--$\la$ propagator must vanish.  Thirdly, the discrete 
symmetries of time-reversal and parity constrain the allowed forms with two 
main consequences: most of the scalar--vector propagators must vanish 
(this is applied in conjunction with the transversality conditions arising 
from enforcing the Lagrange multiplier equations of motion) and the dressing 
functions of the propagator or two-point proper Green's functions must be 
functions of the variables $k_0^2$ and $\vec{k}^2$.

In Table~\ref{tab:w1}, the general decomposition of the propagators 
(collectively denoted by $W$) is presented.  The vector--vector propagators 
are explicitly transverse, with the transverse projector in momentum space 
given by $t_{ij}(\vec{k})=\de_{ij}-k_ik_j/\vec{k}^2$.  It is understood that 
the denominator factors involving both temporal and spatial components 
implicitly carry the appropriate Feynman prescription, i.e.,
\be
\frac{1}{\left(k_0^2-\vec{k}^2\right)}\rightarrow\frac{1}{\left(k_0^2
-\vec{k}^2+\imath0_+\right)},
\ee
such that the integral over the temporal component can be analytically 
continued to Euclidean space ($k_0\rightarrow\imath k_4$).  Supplemented by 
the additional expression for the ghost propagator,
\be
W_{\ov{c}c}^{ab}(k)=-\de^{ab}\frac{\imath D_c}{\vec{k}^2},
\ee
the list is complete.  Each of the dressing functions $D_{\al\ba}$ is a 
function of $k_0^2$ and $\vec{k}^2$ except the ghost which is a function of 
$\vec{k}^2$ only.  The tree-level propagators are given by
\be
D_{AA}=D_{A\pi}=D_{\pi\pi}=D_{\si\si}=D_{\si\phi}=D_{\si\la}=D_c=1,
\;\;\;\;\;\;D_{\phi\phi}=D_{\phi\la}=0.
\ee
The general decomposition of the proper two-point functions (collectively 
denoted by $\G$) is given in Table~\ref{tab:g1}.  The vector--vector 
functions contain longitudinal components and the longitudinal projector is 
written $l_{ij}(\vec{k})=k_ik_j/\vec{k}^2$.  The ghost proper two-point 
function is:
\be
\G_{\ov{c}c}^{ab}(k)=\de^{ab}\imath\G_c\vec{k}^2.
\ee
Again, the dressing functions are functions of $k_0^2$ and $\vec{k}^2$ 
except that for the ghost which is a function of $\vec{k}^2$ only.  At 
tree-level
\be
\G_{AA}=\G_{A\pi}=\G_{\pi\pi}=\G_{\pi\si}=\G_c=1,
\;\;\;\;\;\;\ov{\G}_{AA}=\ov{\G}_{A\pi}=\ov{\G}_{\pi\pi}=\G_{A\si}
=\G_{\si\si}=0.
\ee

The two sets of functions (propagator and two-point proper Green's 
functions) are related via the Legendre transform and we have
\bea
&D_{AA}=\frac{\left(k_0^2-\vec{k}^2\right)\G_{\pi\pi}}{
\left(k_0^2\G_{A\pi}^2-\vec{k}^2\G_{AA}\G_{\pi\pi}\right)},&D_{\si\si}
=\frac{\left(\G_{\pi\pi}+\ov{\G}_{\pi\pi}\right)}{
\G_{\pi\si}^2-\G_{\si\si}\left(\G_{\pi\pi}+\ov{\G}_{\pi\pi}\right)}
,\nonumber\\
&D_{\pi\pi}=\frac{\left(k_0^2-\vec{k}^2\right)\G_{AA}}{
\left(k_0^2\G_{A\pi}^2-\vec{k}^2\G_{AA}\G_{\pi\pi}\right)},&D_{\phi\phi}
=-\frac{\G_{\si\si}}{\G_{\pi\si}^2-\G_{\si\si}
\left(\G_{\pi\pi}+\ov{\G}_{\pi\pi}\right)},\nonumber\\
&D_{A\pi}=\frac{\left(k_0^2-\vec{k}^2\right)\G_{A\pi}}{
\left(k_0^2\G_{A\pi}^2-\vec{k}^2\G_{AA}\G_{\pi\pi}\right)},&D_{\si\phi}
=\frac{\G_{\pi\si}}{\G_{\pi\si}^2-\G_{\si\si}\left(\G_{\pi\pi}
+\ov{\G}_{\pi\pi}\right)},\nonumber\\
&D_c\G_c=1,&\nonumber\\
&0=D_{\si\si}\G_{A\si}-D_{\si\phi}\left(\G_{A\pi}+\ov{\G}_{A\pi}\right)
+D_{\si\la},&\nonumber\\
&0=-D_{\si\phi}\G_{A\si}-D_{\phi\phi}\left(\G_{A\pi}+\ov{\G}_{A\pi}\right)
+D_{\phi\la},&\nonumber\\
&0=\ov{\G}_{AA}-\frac{k_0^2}{\vec{k}^2}\left[D_{\si\la}\G_{A\si}
+D_{\phi\la}\left(\G_{A\pi}+\ov{\G}_{A\pi}\right)\right].&
\label{eq:legtran}
\eea

In addition to the two-point functions we have various three-point 
functions.  The tree-level vertices (three-point proper Green's functions) 
used in this work are given by (all momenta are defined as incoming):
\bea
\G_{\pi\si A ij}^{(0)abc}&=&-gf^{abc}\de_{ij},\nonumber\\
\G_{3A ijk}^{(0)abc}(p_a,p_b,p_c)&=&
-\imath gf^{abc}\left[\de_{ij}(p_a-p_b)_k+\de_{jk}(p_b-p_c)_i
+\de_{ki}(p_c-p_a)_j\right],\nonumber\\
\G_{4A ijkl}^{(0)abcd}&=&-\imath g^2\left\{\de_{ij}\de_{kl}
\left[f^{ace}f^{bde}-f^{ade}f^{cbe}\right]
+\de_{ik}\de_{jl}\left[f^{abe}f^{cde}-f^{ade}f^{bce}\right]
+\de_{il}\de_{jk}\left[f^{ace}f^{dbe}f^{abe}f^{cde}\right]\right\}
,\nonumber\\
\G_{\ov{c}cA i}^{(0)abc}(p_{\ov{c}},p_c,p_A)&=&-\imath gf^{abc}p_{\ov{c}i}
,\nonumber\\
\G_{\phi\si Ai}^{(0)abc}(p_\phi,p_\si,p_A)
&=&\imath p_{\phi j}\G_{\pi\si Aji}^{(0)abc}=-\imath gf^{abc}p_{\phi i}.
\eea

The (Minkowski space) loop integration measure for the integrals entering 
the \DS equations is $(-\dk{\w_M})$ where
\be
\dk{\w_M}=\frac{d\w_0\,d^d\vec{\w}}{(2\pi)^{d+1}}
\ee
and $d=3-2\e$ is the spatial dimension.  In order to preserve the dimension 
of the dressing functions (or, more formally, the action) we must also 
assign a dimension to the coupling through the replacement
\be
g^2\rightarrow g^2\mu^\e
\ee
where $\mu$ is the square of some non-vanishing mass scale (and which will 
be later identified with the renormalization scale).
 
\begin{table}
\begin{tabular}{|c||c|c||c|c||c|c|}\hline
$W$&$A_j$&$\pi_j$&$\si$&$\phi$&$\la$&$\ta$\\
\hline\rule[-2.4ex]{0ex}{5.5ex}
$A_i$&$t_{ij}(\vec{k})\frac{\imath D_{AA}}{(k_0^2-\vec{k}^2)}$&
$t_{ij}(\vec{k})\frac{(-k^0)D_{A\pi}}{(k_0^2-\vec{k}^2)}$&0&0&
$\frac{(-k_i)}{\vec{k}^2}$&0\\
\hline\rule[-2.4ex]{0ex}{5.5ex}
$\pi_i$&$t_{ij}(\vec{k})\frac{k^0D_{A\pi}}{(k_0^2-\vec{k}^2)}$&
$t_{ij}(\vec{k})\frac{\imath\vec{k}^2D_{\pi\pi}}{(k_0^2-\vec{k}^2)}$&0&0&0&
$\frac{(-k_i)}{\vec{k}^2}$\\
\hline\hline\rule[-2.4ex]{0ex}{5.5ex}
$\si$&0&0&$\frac{\imath D_{\si\si}}{\vec{k}^2}$&
$\frac{-\imath D_{\si\phi}}{\vec{k}^2}$&
$\frac{(-k^0)D_{\si\la}}{\vec{k}^2}$&0\\
\hline\rule[-2.4ex]{0ex}{5.5ex}
$\phi$&0&0&$\frac{-\imath D_{\si\phi}}{\vec{k}^2}$&
$\frac{-\imath D_{\phi\phi}}{\vec{k}^2}$&
$\frac{(-k^0)D_{\phi\la}}{\vec{k}^2}$&$\frac{\imath}{\vec{k}^2}$\\
\hline\hline\rule[-2.4ex]{0ex}{5.5ex}
$\la$&$\frac{k_j}{\vec{k}^2}$&0&$\frac{k^0D_{\si\la}}{\vec{k}^2}$&
$\frac{k^0D_{\phi\la}}{\vec{k}^2}$&0&0\\
\hline\rule[-2.4ex]{0ex}{5.5ex}
$\ta$&0&$\frac{k_j}{\vec{k}^2}$&0&$\frac{\imath}{\vec{k}^2}$&0&0\\
\hline
\end{tabular}
\caption{\label{tab:w1}General form of propagators in momentum space.  The 
global color factor $\de^{ab}$ has been extracted.  All unknown functions 
$D_{\al\ba}$ are dimensionless, scalar functions of $k_0^2$ and $\vec{k}^2$.}
\end{table}

\begin{table}
\begin{tabular}{|c||c|c||c||c|c|}\hline
$\G$&$A_j$&$\pi_j$&$\si$&$\;\;\la\;\;$&$\;\;\ta\;\;$\\
\hline\rule[-2.4ex]{0ex}{5.5ex}
$A_i$&$t_{ij}(\vec{k})\imath\vec{k}^2\G_{AA}+\imath k_ik_j\ov{\G}_{AA}$&
$k^0\left(\de_{ij}\G_{A\pi}+l_{ij}(\vec{k})\ov{\G}_{A\pi}\right)$&
$-\imath k^0k_i\G_{A\si}$&$k_i$&0\\
\hline\rule[-2.4ex]{0ex}{5.5ex}
$\pi_i$&$-k^0\left(\de_{ij}\G_{A\pi}+l_{ij}(\vec{k})\ov{\G}_{A\pi}\right)$&
$\imath\de_{ij}\G_{\pi\pi}+\imath l_{ij}(\vec{k})\ov{\G}_{\pi\pi}$&
$k_i\G_{\pi\si}$&
0&$k_i$\\
\hline\hline\rule[-2.4ex]{0ex}{5.5ex}
$\si$&$-\imath k^0k_j\G_{A\si}$&$-k_j\G_{\pi\si}$&
$\imath\vec{k}^2\G_{\si\si}$&0&
0\\\hline\rule[-2.4ex]{0ex}{5.5ex}
$\la$&$-k_j$&0&0&0&0\\\hline\rule[-2.4ex]{0ex}{5.5ex}
$\ta$&0&$-k_j$&0&0&0\\\hline
\end{tabular}
\caption{\label{tab:g1}General form of the proper two-point functions in 
momentum space.   The global color factor $\de^{ab}$ has been extracted.  
All unknown functions $\G_{\al\ba}$ are dimensionless, scalar functions of 
$k_0^2$ and $\vec{k}^2$.}
\end{table}

Although the formalism as presented thus far is written in Minkowski space, 
in order to evaluate the resulting loop integrals it is necessary to 
analytically continue to Euclidean space ($k_0\rightarrow\imath k_4$) and we 
use the notation $k_4$ to denote the temporal component of the Euclidean 
4-momentum such that $k^2=k_4^2+\vec{k}^2$.  The Euclidean integration 
measure is written
\be
\dk{\w}=\frac{d\w_4\,d^d\vec{\w}}{(2\pi)^{d+1}}.
\ee
Since, in this noncovariant setting, the evaluated dressing functions will 
be functions of the two variables $k_4^2$ and $\vec{k}^2$, the Minkowski 
space dressing functions are recovered by analytically continuing back 
with $k_4^2\rightarrow-k_0^2$.  Assuming that the loop integrals can be 
expressed in terms of known analytic functions, the Minkowski space 
dressing functions can be given by simply extending the argument $k_4^2$ 
to negative values and observing contributions generated by continuing 
through singularities.  Physically, such singularities can only occur for 
lightlike momenta (assuming that there are no timelike resonance states) 
just as in linear covariant gauges.  We will use the same notation for the 
Minkowski and Euclidean space functions, since it is clear from the argument 
$k_0^2$ or $k_4^2$ in which space they reside.

Now, the stated purpose of this paper is to derive the one-loop perturbative 
form for the various two-point dressing functions.  In linear covariant 
gauges, the Feynman rules can be easily applied to give directly the 
expressions for propagator functions.  The reason for the relative 
simplicity is that the proper two-point functions are related directly to 
the corresponding propagators via inversion.  This is not so within the 
first order formalism -- since the proper two-point Green's functions and 
the propagators are related effectively by a matrix inversion, one must 
expand the full set of \DS equations for the proper functions in powers of 
the coupling, $g$, and then use the relations \eq{eq:legtran} to construct 
the propagators.  [Actually, a similar situation exists for the quark 
propagator in covariant gauges too, but the matrix inversion only involves 
different contractions of the same equation.]  In the remainder of this 
section, we will thus consider the one-loop perturbative expansion of the 
various \DS equations and write the dressing functions of the proper 
two-point functions in terms of the loop integrals (these integrals will 
be evaluated in the course of the next section).  There are seven such 
equations (derived in \cite{Watson:2006yq}, with the exception of 
$\G_{\si A}$ which is an obvious extension of $\G_{\si\si}$) and they read:
\bea
\G_{\pi Aik}^{ad}(k)&=&-\de^{ad}k^0\de_{ik}
-\int(-\dk{\w_M})\G_{\pi\si Aij}^{(0)abc}(k,-\w,\w-k)W_{\si\ba}^{be}(\w)
\G_{\ba\al Alk}^{efd}(\w,k-\w,-k)W_{\al Alj}^{fc}(\w-k),
\label{eq:dsepa0}\\
\G_c^{ad}(k)&=&\de^{ad}\imath\vec{k}^2
-\int\left(-\dk{\w_M}\right)\G_{\ov{c}cAi}^{(0)abc}(k,-\w,\w-k)W_c^{be}(\w)
W_{A\al ij}^{cf}(k-\w)\G_{\ov{c}c\al j}^{edf}(\w,-k,k-\w),
\label{eq:dsegh0}\\
\G_{\pi\si i}^{ad}(k)&=&\de^{ad}k_i
-\int(-\dk{\w_M})\G_{\pi\si Aij}^{(0)abc}(k,-\w,\w-k)W_{\si\ba}^{be}(\w)
\G_{\ba\al\si l}^{efd}(\w,k-\w,-k)W_{\al Alj}^{fc}(\w-k),
\label{eq:dseps0}\\
\G_{\pi\pi ik}^{ad}(k)&=&\imath\de^{ad}\de_{ik}
-\int(-\dk{\w_M})\G_{\pi\si Aij}^{(0)abc}(k,-\w,\w-k)W_{\si\ba}^{be}(\w)
\G_{\ba\al\pi lk}^{efd}(\w,k-\w,-k)W_{\al Alj}^{fc}(\w-k),
\label{eq:dsepp0}\\
\G_{\si Am}^{ad}(k)&=&
-\int\left(-\dk{\w_M}\right)\G_{\pi\si Aij}^{(0)cab}(\w-k,k,-\w)
W_{A\ba jl}^{be}(\w)\G_{\ba\al A lkm}^{efd}(\w,k-\w,-k)
W_{\al\pi ki}^{fc}(\w-k)\nonumber\\
&&-\int\left(-\dk{\w_M}\right)\G_{\phi\si Ai}^{(0)cab}(\w-k,k,-\w)
W_{A\ba ij}^{be}(\w)\G_{\ba\al A jm}^{efd}(\w,k-\w,-k)W_{\al\phi}^{fc}(\w-k),
\label{eq:dsesa0}\\
\G_{\si\si}^{ad}(k)&=&
-\int\left(-\dk{\w_M}\right)\G_{\pi\si Aij}^{(0)cab}(\w-k,k,-\w)
W_{A\ba jl}^{be}(\w)\G_{\ba\al\si lk}^{efd}(\w,k-\w,-k)
W_{\al\pi ki}^{fc}(\w-k)\nonumber\\
&&-\int\left(-\dk{\w_M}\right)\G_{\phi\si Ai}^{(0)cab}(\w-k,k,-\w)
W_{A\ba ij}^{be}(\w)\G_{\ba\al\si j}^{efd}(\w,k-\w,-k)W_{\al\phi}^{fc}(\w-k),
\label{eq:dsess0}\\
\G_{AAim}^{ae}(k)&=&
\imath\de^{ae}\left[\vec{k}^2\de_{im}-k_ik_m\right]
+\int(-\dk{\w_M})\G_{\ov{c}cA i}^{(0)bca}(\w-k,-\w,k)W_c^{cd}(\w)
\G_{\ov{c}cAm}^{dfe}(\w,k-\w,-k)W_c^{fb}(\w-k)
\nonumber\\&&
-\int(-\dk{\w_M})\G_{\phi\si Ai}^{(0)bca}(\w-k,-\w,k)W_{\si\ba}^{cd}(\w)
\G_{\ba\al Am}^{dfe}(\w,k-\w,-k)W_{\al\phi}^{fb}(\w-k)
\nonumber\\&&
-\int(-\dk{\w_M})\G_{\pi\si Aij}^{(0)bca}(\w-k,-\w,k)W_{\si\ba}^{cd}(\w)
\G_{\ba\al A km}^{dfe}(\w,k-\w,-k)W_{\al\pi kj}^{fb}(\w-k)
\nonumber\\&&
-\frac{1}{2}\int(-\dk{\w_M})\G_{3Akji}^{(0)bca}(\w-k,-\w,k)
W_{A\ba jl}^{cd}(\w)\G_{\ba\al Alnm}^{dfe}(\w,k-\w,-k)W_{\al Ank}^{fb}(\w-k)
\nonumber\\&&
-\frac{1}{6}\int(-\dk{\w_M})(-\dk{v_M})
\G_{4Alkji}^{(0)dcba}(-v,-\w,v+\w-k,k)W_{A\la jn}^{bf}(k-v-\w)
W_{A\ga ko}^{cg}(\w)W_{A\de lp}^{dh}(v)\times
\nonumber\\&&
\G_{\la\ga\de Anopm}^{fghe}(k-\w-v,\w,v,-k)
\nonumber\\&&
+\frac{1}{2}\int(-\dk{\w_M})\G_{4Aimlk}^{(0)aecd}(k,-k,\w,-\w)
W_{AAkl}^{cd}(-\w)
\nonumber\\&&
+\frac{1}{2}\int(-\dk{\w_M})(-\dk{v_M})\G_{4Alkji}^{(0)dcba}(-v,-\w,v+\w-k,k)
W_{A\de ln}^{df}(v)W_{A\ga ko}^{cg}(\w)\G_{\de\ga\la nop}^{fgh}(v,\w,-v-\w)
\times
\nonumber\\&&
W_{\la\mu pq}^{hi}(v+\w)\G_{\mu\nu Aqrm}^{ije}(v+\w,k-v-\w,-k)
W_{\nu Arj}^{jd}(\w+v-k).
\label{eq:dseaa0}
\eea
\begin{table}[t]
\begin{tabular}{|c|c||c|}\hline
equation & integral term & contribution(s) $(\al,\ba,\ldots)$\\\hline
$\G_{\pi A}$, \eq{eq:dsepa0}&1st&$(\al=\pi,\ba=\si)$\\\hline
$\G_c$, \eq{eq:dsegh0}&1st&$(\al=A)$\\\hline
$\G_{\pi\si}$, \eq{eq:dseps0}&1st&$(\al=A,\ba=\phi)$\\\hline
$\G_{\pi\pi}$, \eq{eq:dsepp0}&1st&$(\al=A,\ba=\si)$\\\hline
$\G_{\si A}$, \eq{eq:dsesa0}&1st&$(\al=A,\ba=A)$\\
&2nd&$(\al=\si,\ba=\pi)$\\\hline
$\G_{\si\si}$, \eq{eq:dsess0}&1st&$(\al=A,\ba=\pi)$,$(\al=\pi,\ba=A)$\\
&2nd&---\\\hline
&1st&(explicit)\\
&2nd&$(\al=\si,\ba=\phi)$\\
&3rd&$(\al=\pi,\ba=\si)$\\
$\G_{AA}$, \eq{eq:dseaa0}&4th&$(\al=A,\ba=A)$\\
&5th&---\\
&6th&---\\
&7th&---\\\hline
\end{tabular}
\caption{\label{tab:pert}Contributing integral terms to the one-loop 
perturbative expressions for the \DS equations 
(\ref{eq:dsepa0}--\ref{eq:dseaa0}).}
\end{table}
These equations are all expanded to one-loop, the contributing terms 
tabulated in Table~\ref{tab:pert}.  Introducing some notation for the 
tree-level and one-loop terms, we write
\be
\G_{\al\ba}=\G_{\al\ba}^{(0)}+g^2\G_{\al\ba}^{(1)}
\ee
where the dimensionful parameter $\mu^{\e}$ is included in $\G^{(1)}$.  
After inserting the appropriate tree-level factors, resolving the color 
algebra and analytically continuing to Euclidean space, the one-loop 
expressions read:
\bea
\de_{ij}\G_{A\pi}^{(1)}(k_4^2,\vec{k}^2)+l_{ij}(\vec{k})
\ov{\G}_{A\pi}^{(1)}(k_4^2,\vec{k}^2)&=&
-N_c\mu^\e\int\frac{\dk{\w}\,\w_4}{k_4\w^2(\vec{k}-\vec{\w})^2}
\left[\de_{ik}-\frac{\w_i\w_k}{\vec{\w}^2}\right],
\label{eq:dsepa1}\\
\G_c^{(1)}(\vec{k}^2)&=&
-N_c\mu^\e\int\frac{\dk{\w}}{\w^2(\vec{k}-\vec{\w})^2}
\left[1-\frac{\s{\vec{k}}{\vec{\w}}^2}{\vec{k}^2\vec{\w}^2}\right],
\label{eq:dsegh1}\\
\G_{\pi\si}^{(1)}(k_4^2,\vec{k}^2)&=&
-N_c\mu^\e\int\frac{\dk{\w}}{\w^2(\vec{k}-\vec{\w})^2}
\left[1-\frac{\s{\vec{k}}{\vec{\w}}^2}{\vec{k}^2\vec{\w}^2}\right],
\label{eq:dseps1}\\
\de_{ik}\G_{\pi\pi}^{(1)}(k_4^2,\vec{k}^2)+l_{ik}(\vec{k})
\ov{\G}_{\pi\pi}^{(1)}(k_4^2,\vec{k}^2)&=&
N_c\mu^\e\int\frac{\dk{\w}}{\w^2(\vec{k}-\vec{\w})^2}
\left[\de_{ik}-\frac{\w_i\w_k}{\vec{\w}^2}\right],
\label{eq:dsepp1}\\
\G_{A\si}^{(1)}(k_4^2,\vec{k}^2)&=&
N_c\mu^\e\int\frac{\dk{\w}\,\w_4\s{\vec{k}}{(\vec{k}-2\vec{\w})}}
{k_4\vec{k}^2\w^2(k-\w)^2}\left[d-1-\frac{\vec{k}^2}{\vec{\w}^2}
+\frac{\s{\vec{k}}{(\vec{k}-\vec{\w})}^2}
{\vec{\w}^2(\vec{k}-\vec{\w})^2}\right]\nonumber\\&&
+N_c\mu^\e\int\frac{\dk{\w}\,w_4}{k_4\w^2(\vec{k}-\vec{\w})^2}
\left[1-\frac{\s{\vec{k}}{\vec{\w}}^2}{\vec{k}^2\vec{\w}^2}\right],
\label{eq:dsesa1}\\
\G_{\si\si}^{(1)}(k_4^2,\vec{k}^2)&=&
-N_c\mu^\e\int\frac{\dk{\w}\,\left[(\vec{k}-\vec{\w})^2
-\w_4(\w_4-k_4)\right]}{\vec{k}^2\w^2(k-\w)^2}
\left[d-1-\frac{\vec{k}^2}{\vec{\w}^2}+\frac{\s{\vec{k}}
{(\vec{k}-\vec{\w})}^2}{\vec{\w}^2(\vec{k}-\vec{\w})^2}\right],\nonumber\\
\label{eq:dsess1}\\
t_{im}(\vec{k})\G_{AA}^{(1)}(k_4^2,\vec{k}^2)+l_{im}(\vec{k})
\ov{\G}_{AA}^{(1)}(k_4^2,\vec{k}^2)&=&
N_c\mu^\e\int\frac{\dk{\w}\,\vec{\w}^2}{\vec{k}^2\w^2(\vec{k}-\vec{\w})^2}
\left[\de_{im}-\frac{\w_i\w_m}{\vec{\w}^2}\right]
\nonumber\\&&
-\frac{1}{2}N_c\mu^\e\int\frac{\dk{\w}\,t_{jl}(\vec{\w})t_{nk}
(\vec{k}-\vec{\w})}{\vec{k}^2\w^2(k-\w)^2}\times
\nonumber\\&&
\left[(2\w-k)_i\de_{kj}-2k_k\de_{ij}+2k_j\de_{ik}\right]
\left[(2\w-k)_m\de_{ln}+2k_l\de_{mn}-2k_n\de_{ml}\right].\nonumber\\
\label{eq:dseaa1}
\eea
In the equation, \eq{eq:dseaa0}, for $\G_{AA}$ (the gluon polarization), 
the first two integral terms (which contain the possible energy divergences 
at one-loop) cancel explicitly and this is due to the cancellation of the 
Faddeev-Popov determinant against the determinant arising from the Gau\ss' 
law constraint that is the raison d'\^{e}tre for the first order formalism 
used in this work \cite{Zwanziger:1998ez}.

Let us now discuss in more detail the equations 
(\ref{eq:dsepa1}--\ref{eq:dseaa1}).  Under the transformation 
$\w_4\rightarrow-\w_4$, the right-hand side of Eq.~\eq{eq:dsepa1} and the 
second 
term of the right-hand side of Eq.~\eq{eq:dsesa1} change sign which means 
that 
these integrals must vanish: i.e.,
\be
\G_{A\pi}^{(1)}(k_4^2,\vec{k}^2)=\ov{\G}_{A\pi}^{(1)}(k_4^2,\vec{k}^2)=0
\label{eq:dsepa2}
\ee
and
\be
\G_{A\si}^{(1)}(k_4^2,\vec{k}^2)=
N_c\mu^\e\int\frac{\dk{\w}\,\w_4\s{\vec{k}}{(\vec{k}-2\vec{\w})}}
{k_4\vec{k}^2\w^2(k-\w)^2}\left[d-1-\frac{\vec{k}^2}{\vec{\w}^2}
+\frac{\s{\vec{k}}{(\vec{k}-\vec{\w})}^2}{\vec{\w}^2
(\vec{k}-\vec{\w})^2}\right].
\label{eq:dsesa2}
\ee
Comparing equations \eq{eq:dsegh1} and \eq{eq:dseps1}, we have clearly that
\be
\G_{\pi\si}^{(1)}(k_4^2,\vec{k}^2)=\G_c^{(1)}(\vec{k}^2)
\label{eq:dseps2}
\ee
which is also a consequence of the cancellation of the Faddeev-Popov 
determinant.  The tensor equations, \eq{eq:dsepp1} and \eq{eq:dseaa1}, 
must be decomposed to extract the dressing functions.  We notice however 
that contracting Eq.~\eq{eq:dsepp1} with $l_{ki}(\vec{k})$ gives us the more 
useful combination
\be
\G_{\pi\pi}^{(1)}(k_4^2,\vec{k}^2)+\ov{\G}_{\pi\pi}^{(1)}(k_4^2,\vec{k}^2)
=-\G_c^{(1)}(\vec{k}^2).
\label{eq:dseovpp2}
\ee
On the other hand, contracting Eq.~\eq{eq:dsepp1} with $t_{ki}(\vec{k})$ 
gives
\be
(d-1)\G_{\pi\pi}^{(1)}(k_4^2,\vec{k}^2)=N_c\mu^\e\int\frac{\dk{\w}}
{\w^2(\vec{k}-\vec{\w})^2}\left[d-2+\frac{\s{\vec{k}}{\vec{\w}}^2}
{\vec{k}^2\vec{\w}^2}\right].
\label{eq:dsepp2}
\ee
Turning to Eq.~\eq{eq:dseaa1}, contracting with $l_{mi}(\vec{k})$ gives us
\be
\ov{\G}_{AA}^{(1)}(k_4^2,\vec{k}^2)=N_c\mu^\e\int\frac{\dk{\w}\,
\left(\vec{k}^2\vec{\w}^2-\s{\vec{k}}{\vec{\w}}^2\right)}{\vec{k}^4\w^2
(\vec{k}-\vec{\w})^2}-\frac{N_c}{2}\mu^\e\int\frac{\dk{\w}\,\s{\vec{k}}
{(\vec{k}-2\vec{\w})}^2}{\vec{k}^4\w^2(k-\w)^2}\left[d-1-\frac{\vec{k}^2}
{\vec{\w}^2}+\frac{\s{\vec{k}}{(\vec{k}-\vec{\w})}^2}{\vec{\w}^2(\vec{k}
-\vec{\w})^2}\right],
\label{eq:dseovaa2}
\ee
whereas contraction with $t_{mi}(\vec{k})$ gives 
\bea
(d-1)\G_{AA}^{(1)}(k_4^2,\vec{k}^2)&=&N_c\mu^\e\int\frac{\dk{\w}\,\vec{\w}^2}
{\vec{k}^2\w^2(\vec{k}-\vec{\w})^2}t_{mi}(\vec{k})t_{im}(\vec{\w})
-\frac{N_c}{2}\mu^\e\int\frac{\dk{\w}}{\vec{k}^2\w^2(k-\w)^2}t_{mi}(\vec{k})
t_{jl}(\vec{\w})t_{nk}(\vec{k}-\vec{\w})\times
\nonumber\\&&
\left[(2\w-k)_i\de_{kj}-2k_k\de_{ij}+2k_j\de_{ik}\right]
\left[(2\w-k)_m\de_{ln}+2k_l\de_{mn}-2k_n\de_{ml}\right].
\label{eq:dseaa2}
\eea
Expanding the transverse projectors of Eq.~\eq{eq:dseaa2}, contracting 
indices and where possible canceling denominator factors, we get
\bea
(d-1)\G_{AA}^{(1)}(k_4^2,\vec{k}^2)&=&N_c\mu^\e\int\frac{\dk{\w}}
{\vec{k}^2\w^2(\vec{k}-\vec{\w})^2}\left[(d-2)\vec{\w}^2
+\frac{\s{\vec{k}}{\vec{\w}}^2}{\vec{k}^2}\right]\nonumber\\
&&+N_c\mu^\e\int\frac{\dk{\w}}{\vec{k}^2\w^2(k-\w)^2}
\left\{(6-4d)\vec{k}^2+2(d-1)\frac{\s{\vec{k}}{\vec{\w}}^2}{\vec{k}^2}
+2(1-d)\vec{\w}^2\right.\nonumber\\
&&\left.+\frac{1}{\vec{\w}^2}\left[-\frac{1}{4}\vec{k}^4-\ha\vec{k}^2
\s{\vec{k}}{\vec{\w}}+(4d-5)\s{\vec{k}}{\vec{\w}}^2-2\frac{\s{\vec{k}}
{\vec{\w}}^3}{\vec{k}^2}\right]+\frac{1}{8}\frac{\vec{k}^6}
{\vec{\w}^2(\vec{k}-\vec{\w})^2}\right\}.
\eea
In order to reduce the number of integrals that we need to compute, we use 
the identities
\bea
\s{\vec{k}}{\vec{\w}}&=&1/2(k^2+\w^2-(k-\w)^2)-k_4\w_4,\nonumber\\
\w_4^2&=&\w^2-\vec{\w}^2\nonumber
\eea
to get
\bea
(d-1)\G_{AA}^{(1)}(k_4^2,\vec{k}^2)&=&\frac{N_c}{8}\mu^\e\int\frac{\dk{\w}\,
\vec{k}^4}{\w^2(k-\w)^2\vec{\w}^2(\vec{k}-\vec{\w})^2}
+\left[1+(10-8d)\frac{k^2}{\vec{k}^2}+3\frac{k^4}{\vec{k}^4}\right]
\frac{N_c}{2}\mu^\e\int\frac{\dk{\w}\,k_4\w_4}{\w^2(k-\w)^2\vec{\w}^2}
\nonumber\\&&
+\left[(4d-5)\frac{k^4}{\vec{k}^2}-k^2-\vec{k}^2-\frac{k^6}{\vec{k}^4}\right]
\frac{N_c}{4}\mu^\e\int\frac{\dk{\w}}{\w^2(k-\w)^2\vec{\w}^2}
\nonumber\\&&
+N_c\mu^\e\int\frac{\dk{\w}}{\w^2(k-\w)^2}\left[1+(3-4d)\frac{k^2}{\vec{k}^2}
+2\frac{k^4}{\vec{k}^4}+2\frac{\s{\vec{k}}{\vec{\w}}}{\vec{k}^2}
\left(\frac{k^2}{\vec{k}^2}-1\right)+2(1-d)\left(\frac{\vec{\w}^2}{\vec{k}^2}
-\frac{\s{\vec{k}}{\vec{\w}}^2}{\vec{k}^4}\right)\right]
\nonumber\\&&
+N_c\mu^\e\int\frac{\dk{\w}}{\w^2(\vec{k}-\vec{\w})^2}\left[-\frac{1}{4}
+\left(3d-\frac{17}{4}\right)\frac{k^2}{\vec{k}^2}-\frac{5}{4}\frac{k^4}
{\vec{k}^4}+\frac{\s{\vec{k}}{\vec{\w}}}{\vec{k}^2}\left(\frac{7}{2}-2d
+\frac{3}{2}\frac{k^2}{\vec{k}^2}\right)
+(d-2)\frac{\vec{\w}^2}{\vec{k}^2}\right].\nonumber\\
\label{eq:dseaa3}
\eea
We must also deal with the integrals for the $\G_{\si\si}$, $\G_{A\si}$ 
and $\ov{\G}_{AA}$ Green's functions, given by equations (\ref{eq:dsess1}), 
(\ref{eq:dsesa2}) and (\ref{eq:dseovaa2}).  It turns out that all 
these expressions are related as we will now show.  Starting with 
$\G_{A\si}$, Eq.~\eq{eq:dsesa2}, we can easily show that
\be
\G_{A\si}^{(1)}(k_4^2,\vec{k}^2)=
(d-1)N_c\mu^\e\int\frac{\dk{\w}\,\w_4\s{\vec{k}}{(\vec{k}-2\vec{\w})}}
{k_4\vec{k}^2\w^2(k-\w)^2}
+N_c\mu^\e\int\frac{\dk{\w}\,(2\w_4-k_4)\s{\vec{k}}{\vec{\w}}^2}
{k_4\vec{k}^2\w^2(k-\w)^2\vec{\w}^2}.
\label{eq:dsesa3}
\ee
The second integral expression will be reduced below but it is convenient 
for now to leave it in its present form.  Now consider Eq.~\eq{eq:dsess1}.  
If we first recall that the last factor was originally written in symmetric 
form
\be
t_{ij}(\vec{\w})t_{ji}(\vec{k}-\vec{\w})=
\left[d-1-\frac{\vec{k}^2}{\vec{\w}^2}+\frac{\s{\vec{k}}{(\vec{k}
-\vec{\w})}^2}{\vec{\w}^2(\vec{k}-\vec{\w})^2}\right]
\ee
then it is easy to show that
\be
\G_{\si\si}^{(1)}(k_4^2,\vec{k}^2)=
N_c\mu^\e\int\frac{\dk{\w}\,(k_4-2\w_4)}{k_4\vec{k}^2\w^2(k-\w)^2}
\left[(d-1)\vec{\w}^2-\vec{k}^2\right]
+N_c\mu^\e\int\frac{\dk{\w}\,(2\w_4-k_4)\s{\vec{k}}{\vec{\w}}^2}
{k_4\vec{k}^2\w^2(k-\w)^2\vec{\w}^2}.
\ee
We immediately recognize the latter factor as occurring in 
Eq.~\eq{eq:dsesa3}.  
As for the former, firstly we have as a general result that
\be
\int\frac{\dk{\w}\,(k_4-2\w_4)}{\w^2(k-\w)^2}=0.
\ee
It can also be shown (using the results of Appendix~\ref{app:int0}) that
\be
\int\frac{\dk{\w}\,(k_4-2\w_4)\vec{\w}^2}{k_4\vec{k}^2\w^2(k-\w)^2}
=\int\frac{\dk{\w}\,\w_4\s{\vec{k}}{(\vec{k}-2\vec{\w})}}
{k_4\vec{k}^2\w^2(k-\w)^2}.
\ee
Thus we see that
\be
\G_{\si\si}^{(1)}(k_4^2,\vec{k}^2)=\G_{A\si}^{(1)}(k_4^2,\vec{k}^2).
\label{eq:dsess3}
\ee
Turning now to Eq.~\eq{eq:dseovaa2}, it is easy to show that
\be
\ov{\G}_{AA}^{(1)}(k_4^2,\vec{k}^2)=-N_c\mu^\e\frac{(d-1)}{2}
\int\frac{\dk{\w}\,\s{\vec{k}}{(\vec{k}-2\vec{\w}})^2}{\vec{k}^4\w^2(k-\w)^2}
-N_c\mu^\e\int\frac{\dk{\w}\,k_4(2\w_4-k_4)\s{\vec{k}}{\vec{\w}}^2}
{\vec{k}^4\w^2(k-\w)^2\vec{\w}^2}
\ee
and further manipulating, we get that
\be
\ov{\G}_{AA}^{(1)}(k_4^2,\vec{k}^2)=-\frac{k_4^2}{\vec{k}^2}
\G_{A\si}^{(1)}(k_4^2,\vec{k}^2).
\label{eq:dseovaa3}
\ee
Let us finally reduce the expression for $\G_{A\si}$, Eq.~\eq{eq:dsesa3}, to 
the set of most 
basic integrals as we did for $\G_{AA}$.  Using the same techniques as 
before, we have that
\bea
\G_{A\si}^{(1)}(k_4^2,\vec{k}^2)&=&\left[\frac{k^2}{\vec{k}^2}
+\ha\frac{k^4}{k_4^2\vec{k}^2}\right]N_c\mu^\e\int\frac{\dk{\w}\,k_4\w_4}
{\w^2(k-\w)^2\vec{\w}^2}-\frac{1}{4}\frac{k^4}{\vec{k}^2}N_c\mu^\e
\int\frac{\dk{\w}}{\w^2(k-\w)^2\vec{\w}^2}
\nonumber\\&&
+N_c\mu^\e\int\frac{\dk{\w}}{\w^2(k-\w)^2}\left[(d-1)\frac{\w_4}{k_4}-1
+2\frac{k^2}{\vec{k}^2}-2(d-1)\frac{\w_4\s{\vec{k}}{\vec{\w}}}{k_4\vec{k}^2}
+2\frac{\s{\vec{k}}{\vec{\w}}}{\vec{k}^2}\right]
\nonumber\\&&
+N_c\mu^\e\int\frac{\dk{\w}}{\w^2(\vec{k}-\vec{\w})^2}\left[-1
-\frac{5}{4}\frac{k^2}{\vec{k}^2}
+\frac{3}{2}\frac{\s{\vec{k}}{\vec{\w}}}{\vec{k}^2}\right].
\label{eq:dsesa4}
\eea

Before proceeding, let us briefly summarize the results of this section.  We 
have written the various one-loop, two-point proper Green's functions in 
terms of the most simple collection of integrals.  $\G_{A\pi}^{(1)}$ and 
$\ov{\G}_{A\pi}^{(1)}$ are trivial, Eq.~\eq{eq:dsepa2}.  We must calculate 
$\G_c^{(1)}$, Eq.~\eq{eq:dsegh1}, which will also yield $\G_{\pi\si}^{(1)}$ 
via Eq.~\eq{eq:dseps2}.  With $\G_{\pi\pi}^{(1)}$ calculated using 
Eq.~\eq{eq:dsepp2}, we also get $\ov{\G}_{\pi\pi}^{(1)}$ from 
Eq.~\eq{eq:dseovpp2}.  
$\G_{A\si}^{(1)}$ is calculated using Eq.~\eq{eq:dsesa4} which then gives us 
$\G_{\si\si}^{(1)}$ and $\ov{\G}_{AA}^{(1)}$ from relations \eq{eq:dsess3} 
and \eq{eq:dseovaa3}, respectively.  Finally, we have to calculate 
$\G_{AA}^{(1)}$ using Eq.~\eq{eq:dseaa3}.  The necessary integrals will be 
derived in the next section.

\section{Non-covariant Loop Integrals}
\setcounter{equation}{0}
In this section we consider the nontrivial loop integrals that arise in 
this study (i.e., massless one-loop two-point integrals).  Quite generally, 
the integrals that arise within Coulomb gauge perturbation theory can be 
classified into two categories -- those that can be evaluated using standard 
techniques such as Feynman parametrization or Schwinger parameters (and 
which are detailed in Appendix~\ref{app:int0}) and those that cannot.  The 
latter category clearly requires a different approach and to this effect we 
derive a technique based on differential equations and integration by parts 
[IBP] suitable for the non-covariant setting here.  We consider the three 
integrals:
\bea
A(k_4^2,\vec{k}^2)&=&\int\frac{\dk{\w}}{\w^2(k-\w)^2\vec{\w}^2},
\label{eq:adef}\\
A^4(k_4^2,\vec{k}^2)&=&\int\frac{\dk{\w}\,\w_4}{\w^2(k-\w)^2\vec{\w}^2},
\label{eq:a4def}\\
B(k_4^2,\vec{k}^2)&=&\int\frac{\dk{\w}}{\w^2(k-\w)^2\vec{\w}^2
(\vec{k}-\vec{\w})^2}.\label{eq:bdef}
\eea
It is convenient to introduce the following notation: $x=k_4^2$, 
$y=\vec{k}^2$, $z=x/y$, $v=y/x$.  We find that the above integrals can be 
written in the form
\bea
A(x,y)&=&\frac{(x+y)^{-1-\e}}{(4\pi)^{2-\e}}f_a(z),\label{eq:afn}\\
A^4(x,y)=k_4\ov{A}(x,y)&=&k_4\frac{(x+y)^{-1-\e}}{(4\pi)^{2-\e}}f_4(z)
\label{eq:a4fn},\\
B(x,y)&=&\frac{y^{-2}(x+y)^{-\e}}{(4\pi)^{2-\e}}f_b(z)\label{eq:bfn}
\eea
where the $f_i(z)$ are functions of $z$ and which may contain an ultraviolet 
divergence in the form of a simple pole as $\e\rightarrow0$.  The functions 
$f_i(z)$ can (and will) be written down in analytic form for 
$\e\rightarrow0$, but it turns out to be more useful to write some of the 
various parts in terms of an integral representation and also as asymptotic 
series in $z$ or $v$ which are more amenable to eventual numerical 
evaluation.

\subsection{Derivation of the Differential Equations}
To begin, let us derive the differential equations that these integrals 
obey.  Consider the general integral ($n=0,1$)
\be
I^n(k_4^2,\vec{k}^2)=\int\frac{\dk{\w}\,\w_4^n}{\w^2(k-\w)^2\vec{\w}^2}.
\ee
Since $I^n$ is a function of two variables, there are two first derivatives:
\bea
k_4\frac{\pd I^n}{\pd k_4}&=&\int\frac{\dk{\w}\,\w_4^n}{\w^2(k-\w)^2
\vec{\w}^2}\left\{-2\frac{k_4(k_4-\w_4)}{(k-\w)^2}\right\},
\label{eq:ade0}\\
k_k\frac{\pd I^n}{\pd k_k}&=&\int\frac{\dk{\w}\,\w_4^n}{\w^2(k-\w)^2
\vec{\w}^2}\left\{-2\frac{\s{\vec{k}}{(\vec{k}-\vec{\w})}}{(k-\w)^2}\right\}.
\eea
Now, there are also two integration by parts identities:
\bea
0=\int\dk{\w}\frac{\pd}{\pd\w_4}\frac{\w_4^{n+1}}{\w^2(k-\w)^2\vec{\w}^2}
&=&\int\frac{\dk{\w}\,\w_4^n}{\w^2(k-\w)^2\vec{\w}^2}
\left\{n+1-2\frac{\w_4^2}{\w^2}-2\frac{\w_4(\w_4-k_4)}{(k-\w)^2}\right\},
\label{eq:ibp1}\\
0=\int\dk{\w}\frac{\pd}{\pd\w_i}\frac{\w_i\w_4^n}{\w^2(k-\w)^2\vec{\w}^2}
&=&\int\frac{\dk{\w}\,\w_4^n}{\w^2(k-\w)^2\vec{\w}^2}
\left\{d-2-2\frac{\vec{\w}^2}{\w^2}-2\frac{\s{\vec{\w}}{(\vec{\w}-\vec{k})}}
{(k-\w)^2}\right\}.
\eea
Adding these two expressions gives
\be
0=\int\frac{\dk{\w}\,\w_4^n}{\w^2(k-\w)^2\vec{\w}^2}
\left\{d+n-5+2\frac{\s{k}{(k-\w)}}{(k-\w)^2}\right\}
\label{eq:ibp0}
\ee
from which we have the important identity
\be
k_4\frac{\pd I^n}{\pd k_4}+k_k\frac{\pd I^n}{\pd k_k}=(d+n-5)I^n.
\label{eq:de0}
\ee
Expanding the numerator factor, Eq.~\eq{eq:ibp0} can be rewritten
\be
0=\int\frac{\dk{\w}\,\w_4^n}{\w^2(k-\w)^2\vec{\w}^2}
\left\{d+n-4+\frac{k^2-\w^2}{(k-\w)^2}\right\}.
\label{eq:ibp3}
\ee
Similarly, Eq.~\eq{eq:ibp1} becomes
\be
0=\int\frac{\dk{\w}\,\w_4^n}{\w^2(k-\w)^2\vec{\w}^2}
\left\{n-1+2\frac{\vec{\w}^2}{\w^2}
+\frac{\left[-(k_4-\w_4)^2+k_4^2-\w^2+\vec{\w}^2\right]}{(k-\w)^2}\right\}.
\ee
We can now rewrite Eq.~\eq{eq:ade0} as
\be
k_4\frac{\pd I^n}{\pd k_4}=
\int\frac{\dk{\w}\,\w_4^n}{\w^2(k-\w)^2\vec{\w}^2}
\left\{1-n+\frac{k_4^2}{k^2}\left[d+n-4\right]-2\frac{\vec{\w}^2}{\w^2}
+2\frac{\vec{k}^2}{k^2}\frac{\w^2}{(k-\w)^2}
-2\frac{\vec{\w}^2}{(k-\w)^2}\right\}
\ee
and so we arrive at the temporal differential equations for $A$ and $A^4$:
\bea
k_4\frac{\pd A}{\pd k_4}&=&\left[1+2(d-4)\frac{k_4^2}{k^2}\right]A
+2\frac{\vec{k}^2}{k^2}\int\frac{\dk{\w}}{(k-\w)^4\vec{\w}^2}
-2\int\frac{\dk{\w}}{\w^4(k-\w)^2}-2\int\frac{\dk{\w}}{\w^2(k-\w)^4},\\
k_4\frac{\pd A^4}{\pd k_4}&=&\left[2(d-3)\frac{k_4^2}{k^2}\right]A^4
+2\frac{\vec{k}^2}{k^2}\int\frac{\dk{\w}\,\w_4}{(k-\w)^4\vec{\w}^2}
-2\int\frac{\dk{\w}\,\w_4}{\w^4(k-\w)^2}
-2\int\frac{\dk{\w}\,\w_4}{\w^2(k-\w)^4}.
\eea
The differential equations involving the spatial components are subsequently 
given using Eq.~\eq{eq:de0}.

Now let us consider integrals of the form $B$.  Again, we have two IBP 
identities:
\bea
0=\int\dk{\w}\frac{\pd}{\pd\w_4}\frac{\w_4}{\w^2(k-\w)^2\vec{\w}^2(\vec{k}
-2\vec{\w})^2}&=&
\int\frac{\dk{\w}}{\w^2(k-\w)^2\vec{\w}^2(\vec{k}-2\vec{\w})^2}
\left\{1-2\frac{\w_4^2}{\w^2}-2\frac{\w_4(\w_4-k_4)}{(k-\w)^2}\right\},
\label{eq:ibp2}\\
0=\int\dk{\w}\frac{\pd}{\pd\w_i}\frac{\w_i}{\w^2(k-\w)^2\vec{\w}^2
(\vec{k}-2\vec{\w})^2}&=&
\int\frac{\dk{\w}}{\w^2(k-\w)^2\vec{\w}^2(\vec{k}-2\vec{\w})^2}
\left\{d-2-2\frac{\vec{\w}^2}{\w^2}
-2\frac{\s{\vec{\w}}{(\vec{\w}-\vec{k})}}{(k-\w)^2}
-2\frac{\s{\vec{\w}}{(\vec{\w}-\vec{k})}}{(\vec{k}-\vec{\w})^2}\right\},
\nonumber\\
\eea
and by adding the two we see that
\be
0=\int\frac{\dk{\w}}{\w^2(k-\w)^2\vec{\w}^2(\vec{k}-2\vec{\w})^2}
\left\{d-7+2\frac{\s{k}{(k-\w)}}{(k-\w)^2}
+2\frac{\s{\vec{k}}{(\vec{k}-\vec{\w})}}{(\vec{k}-\vec{\w})^2}\right\}
\ee
from which we have the important identity
\be
k_4\frac{\pd B}{\pd k_4}+k_k\frac{\pd B}{\pd k_k}=(d-7)B.
\label{eq:de1}
\ee
Rewriting the definition of $B$ as
\be
2B=2\int\frac{\dk{\w}\,w_4^2}{\w^4(k-\w)^2\vec{\w}^2(\vec{k}-2\vec{\w})^2}
+2\int\frac{\dk{\w}}{\w^4(k-\w)^2(\vec{k}-\vec{\w})^2}
\ee
and using Eq.~\eq{eq:ibp2} gives
\bea
B-2\int\frac{\dk{\w}}{\w^2(k-\w)^4\vec{\w}^2}
&=&\int\frac{\dk{\w}}{\w^2(k-\w)^2\vec{\w}^2(\vec{k}-2\vec{\w})^2}
\left\{-2\frac{\w_4(\w_4-k_4)}{(k-\w)^2}\right\}\nonumber\\
&=&\int\frac{\dk{\w}}{\w^2(k-\w)^2\vec{\w}^2(\vec{k}-2\vec{\w})^2}
\left\{-2\frac{(\w_4-k_4)^2}{(k-\w)^2}+2\frac{k_4(k_4-\w)}{(k-\w)^2}\right\}
\nonumber\\
&=&-2B+2\int\frac{\dk{\w}}{\w^2(k-\w)^4\vec{\w}^2}-k_4\frac{\pd B}{\pd k_4}.
\eea
However, with Eq.~\eq{eq:ibp3} we obtain the temporal differential equation 
for 
$B$:
\be
k_4\frac{\pd B}{\pd k_4}=-3B+\frac{4(4-d)}{k^2}A+\frac{4}{k^2}
\int\frac{\dk{\w}}{\w^4(\vec{k}-\vec{\w})^4}
\ee
and the spatial equation is given via Eq.~\eq{eq:de1}.

The differential equations are best written by evaluating the standard 
integrals in terms of 
$\e$ (see Appendix~\ref{app:int0}) and with the notation $x$ and $y$ for 
the temporal and spatial momentum components.  The result is 
($A^4=k_4\ov{A}$):
\bea
2x\frac{\pd A}{\pd x}&=&\left[1-(2+4\e)\frac{x}{x+y}\right]A
-2\frac{y^{-\e}}{x+y}X+4(x+y)^{-1-\e}Y,\label{eq:atde0}\\
2x\frac{\pd\ov{A}}{\pd x}&=&\left[-1-4\e\frac{x}{x+y}\right]\ov{A}
-2\frac{y^{-\e}}{x+y}X+2(x+y)^{-1-\e}Y,\label{eq:a4tde0}\\
2x\frac{\pd B}{\pd x}&=&-3B+(4+8\e)(x+y)^{-1}A-4\frac{y^{-1-\e}}{x+y}X,
\label{eq:btde0}
\eea
where
\be
X=\frac{1}{(4\pi)^{2-\e}}\G(\e)\G(1-\e)\frac{\G(1/2-\e)}{\G(1/2-2\e)},
\;\;\;\;Y=\frac{1}{(4\pi)^{2-\e}}\G(\e)\G(1-\e)\frac{\G(1-\e)}{\G(1-2\e)}.
\ee
The spatial differential equations subsequently read:
\bea
2y\frac{\pd A}{\pd y}&=&\left[-3-2\e+(2+4\e)\frac{x}{x+y}\right]A
+2\frac{y^{-\e}}{x+y}X-4(x+y)^{-1-\e}Y,\label{eq:asde0}\\
2y\frac{\pd\ov{A}}{\pd y}&=&\left[-1-2\e+4\e\frac{x}{x+y}\right]\ov{A}
+2\frac{y^{-\e}}{x+y}X-2(x+y)^{-1-\e}Y,\label{eq:a4sde0}\\
2y\frac{\pd B}{\pd y}&=&(-1-2\e)B-(4+8\e)(x+y)^{-1}A
+4\frac{y^{-1-\e}}{x+y}X.\label{eq:bsde0}
\eea
\subsection{Solving the Differential Equations}
The differential equations (\ref{eq:atde0}-\ref{eq:btde0},
\ref{eq:asde0}-\ref{eq:bsde0}) have a rather special structure that allows 
for a 
relatively simple solution.  Taking $A$ to start (the other two follow the 
same pattern), we can write
\be
A(x,y)=F_A(x,y)G_A(x,y)
\ee
such that
\bea
2x\frac{\pd F_A(x,y)}{\pd x}&=&\left[1-(2+4\e)\frac{x}{x+y}\right]F_A(x,y),\\
2y\frac{\pd F_A(x,y)}{\pd y}&=&\left[-3-2\e+(2+4\e)\frac{x}{x+y}\right]
F_A(x,y),\\
F_A(x,y)2x\frac{\pd G_A(x,y)}{\pd x}
&=&-2\frac{y^{-\e}}{x+y}X+4(x+y)^{-1-\e}Y,\\
F_A(x,y)2y\frac{\pd G_A(x,y)}{\pd y}
&=&2\frac{y^{-\e}}{x+y}X-4(x+y)^{-1-\e}Y.
\eea
The solution for the first pair of equations can be written down without 
difficulty and is
\be
F_A(x,y)=x^{1/2}y^{-1/2+\e}(x+y)^{-1-2\e}+{\cal C}.
\ee
By inspection, the constant ${\cal C}=0$ since the function $F_A(x,y)$ has 
the dimension $[y]^{-\e}$ (i.e., the function $F_A$ carries the dimension of 
the integral $A$).  The remaining differential equations for $G_A$ now read
\bea
x^{1/2}y^{-1/2+\e}(x+y)^{-1-2\e}2x\frac{\pd G_A(x,y)}{\pd x}
&=&-2\frac{y^{-\e}}{x+y}X+4(x+y)^{-1-\e}Y,\\
x^{1/2}y^{-1/2+\e}(x+y)^{-1-2\e}2y\frac{\pd G_A(x,y)}{\pd y}
&=&2\frac{y^{-\e}}{x+y}X-4(x+y)^{-1-\e}Y.
\eea
Because the derivatives with respect to $x$ and $y$ are distinguished only 
by a minus sign, the function $G_A$ must be a dimensionless function of the 
dimensionless ratio $v=y/x$ or (equivalently) $z=x/y$.  We choose to express 
$G_A$ as a function of $v$ to avoid singularities and the two partial 
differential equations collapse into a single first order ordinary 
differential equation:
\be
\frac{dG_A(v)}{dv}=v^{-1/2-2\e}(1+v)^{2\e}X-2v^{-1/2-\e}(1+v)^{\e}Y.
\ee
Using the integral representation of the hypergeometric function 
\cite{abram}:
\be
\int_0^v dt\,t^{b-1}(1+t)^{-a}=\frac{1}{b}\frac{v^b}{(1+v)^a}
{}_2F_1\left(a,1;1+b;\frac{v}{1+v}\right),
\ee
gives the solution
\bea
G_A(v)-G_A(0)&=&\frac{1}{1/2-2\e}\frac{v^{1/2-2\e}}{(1+v)^{-2\e}}X
{}_2F_1\left(-2\e,1;3/2-2\e;\frac{v}{1+v}\right)\nonumber\\
&&-\frac{2}{1/2-\e}\frac{v^{1/2-\e}}{(1+v)^{-\e}}X
{}_2F_1\left(-\e,1;3/2-\e;\frac{v}{1+v}\right).
\eea
Let us now show that the constant $G_A(0)=0$.  The general solution for 
$A(x,y)$ as $y\rightarrow0$ would have a term 
\be
F_A(x,y\rightarrow0)G_A(0)\sim\lim_{y\rightarrow0}x^{1/2}y^{-1/2+\e}
\ee
but in the original partial differential equation \eq{eq:atde0}, which must 
still be defined for all values of $y$ (although the coefficients may be 
singular) there is no such term and the constant $G_A(0)$ must vanish.  We 
thus have the solution
\bea
\lefteqn{A(x,y)=\frac{1}{(x+y)^{1+\e}}\times}\nonumber\\&&
\left\{\frac{X}{1/2-2\e}\left(\frac{v}{1+v}\right)^{-e}
{}_2F_1\left(-2\e,1;3/2-2\e;\frac{v}{1+v}\right)-\frac{2Y}{1/2-\e}
{}_2F_1\left(-\e,1;3/2-\e;\frac{v}{1+v}\right)\right\}.
\eea
As it stands, the solution is not of much use since the hypergeometric 
functions are somewhat cumbersome.  However, we are primarily interested in 
the solution as $\e\rightarrow0$.  In this case
\be
{}_2F_1\left(-\e,1;3/2-\e;\frac{v}{1+v}\right)=1+\e f(v)+{\cal O}(\e^2)
\ee
and upon expanding $A$ in powers of $\e$ the functions $f(v)$ happen to 
cancel such that
\be
A(x,y)=\frac{(x+y)^{-1-\e}}{(4\pi)^{2-\e}}\left\{-2\left(\frac{1}{\e}
-\ga\right)-4\ln{2}+2\ln{\left(\frac{x+y}{y}\right)}+{\cal O}(\e)\right\}.
\label{eq:asol}
\ee
A few remarks are in order here.  The overall dimension of the integral $A$ 
can be written in terms of the covariant factor $x+y$ as for the integrals 
in linear covariant gauges and result in the standard logarithmic factor, 
singular on the light-cone and with a branch-cut extending into the timelike 
region.  The ultraviolet divergence characterized by the $1/\e$ term is 
similarly covariant.  The noncovariant component is logarithmically 
singular at $y=0$.  The $x$-dependence of the noncovariant component, 
however, has the logarithmic singularity at $x+y=0$.  This is actually 
rather crucial since otherwise it would be difficult to justify the 
analytic continuation between Euclidean and Minkowski space without further 
cancellations in forming the two-point functions.  Lastly, it is remarkable 
that the integral $A$ which defies standard evaluation techniques reduces 
merely to a combination of logarithms.  Unfortunately, this simplicity will 
not be present in the integrals $A^4$ and $B$.

Let us now turn to the function $\ov{A}$ governed by equations 
\eq{eq:a4tde0} and 
\eq{eq:a4sde0}.  Writing $\ov{A}(x,y)=F_{\ov{A}}G_{\ov{A}}$ as before leads 
to four partial differential equations:
\bea
2x\frac{\pd F_{\ov{A}}(x,y)}{\pd x}&=&\left[1-4\e\frac{x}{x+y}\right]
F_{\ov{A}}(x,y),\\
2y\frac{\pd F_{\ov{A}}(x,y)}{\pd y}&=&\left[-1-2\e+4\e\frac{x}{x+y}\right]
F_{\ov{A}}(x,y),\\
F_{\ov{A}}(x,y)2x\frac{\pd G_{\ov{A}}(x,y)}{\pd x}&=&-2\frac{y^{-\e}}{x+y}X
+2(x+y)^{-1-\e}Y,\\
F_{\ov{A}}(x,y)2y\frac{\pd G_{\ov{A}}(x,y)}{\pd y}&=&2\frac{y^{-\e}}{x+y}X
-2(x+y)^{-1-\e}Y.
\eea
The solution to the first pair is
\be
F_{\ov{A}}(x,y)=x^{-1/2}y^{-1/2+\e}(x+y)^{-2\e},
\ee
the constant vanishing on dimensional grounds as previously.  The latter 
two differential equations now read
\bea
x^{-1/2}y^{-1/2+\e}(x+y)^{-2\e}2x\frac{\pd G_{\ov{A}}(x,y)}{\pd x}
&=&-2\frac{y^{-\e}}{x+y}X+2(x+y)^{-1-\e}Y,\\
x^{-1/2}y^{-1/2+\e}(x+y)^{-2\e}2y\frac{\pd G_{\ov{A}}(x,y)}{\pd y}
&=&2\frac{y^{-\e}}{x+y}X-2(x+y)^{-1-\e}Y.
\eea
Again, as before the relative minus sign means that $G_{\ov{A}}$ is a 
function of the ratio, but this time it is useful to use both $v$ and $z$, 
the reason becoming clear shortly.  In terms of $v$, we have that
\be
\frac{dG_{\ov{A}}(v)}{dv}=v^{-1/2-2\e}(1+v)^{-1+2\e}X
-v^{-1/2-\e}(1+v)^{-1+\e}Y
\ee
whose solution is
\be
G_{\ov{A}}(v)=\frac{X}{1/2-2\e}\frac{v^{1/2-2\e}}{(1+v)^{1-2\e}}
{}_2F_1\left(1-2\e,1;3/2-2\e;\frac{v}{1+v}\right)-\frac{Y}{1/2-\e}
\frac{v^{1/2-\e}}{(1+v)^{1-\e}}
{}_2F_1\left(1-\e,1;3/2-\e;\frac{v}{1+v}\right),
\ee
whereas in terms of $z$ we get that
\be
\frac{dG_{\ov{A}}(z)}{dz}=-z^{-1/2}(1+z)^{-1+2\e}X+z^{-1/2}(1+z)^{-1+\e}Y
\ee
whose solution is
\be
G_{\ov{A}}(z)=-\frac{2X}{1/2-2\e}\frac{z^{1/2}}{(1+z)^{1-2\e}}
{}_2F_1\left(1-2\e,1;3/2;\frac{z}{1+z}\right)+\frac{2Y}{1/2-\e}
\frac{z^{1/2}}{(1+z)^{1-\e}}{}_2F_1\left(1-\e,1;3/2;\frac{z}{1+z}\right).
\ee
The general constant terms in both solutions vanish as before.  We can now 
write the result for the full integral $A^4(x,y)=k_4\ov{A}$ in two ways:
\bea
\lefteqn{A^4(x,y)=\frac{k_4}{(x+y)^{1+\e}}\times}\nonumber\\&&
\left\{\frac{X}{1/2-2\e}\left(\frac{v}{1+v}\right)^{-e}
{}_2F_1\left(1-2\e,1;3/2-2\e;\frac{v}{1+v}\right)-\frac{Y}{1/2-\e}
{}_2F_1\left(1-\e,1;3/2-\e;\frac{v}{1+v}\right)\right\},\\&&
A^4(x,y)=\frac{k_4}{(x+y)^{1+\e}}\left\{-2X(1+z)^{\e}
{}_2F_1\left(1-2\e,1;3/2;\frac{z}{1+z}\right)
+2Y{}_2F_1\left(1-\e,1;3/2;\frac{z}{1+z}\right)\right\}.
\eea
Now, in contradistinction to the integral $A$, the expansion of the 
hypergeometric functions in powers of $\e$ does not yield a simple 
result -- quite the contrary.  However, one can expand in powers of either 
$v$ or $z$ and then take the limit $\e\rightarrow0$ which will prove useful 
in later numerical evaluation.  Actually, in the case of the expansion with 
$v$, one must first collect together terms involving $\ln{v/(1+v)}$ since 
these factors do not have an expansion around $v=0$.  The results are 
ultraviolet finite and are:
\bea
A^4(x,y)&\stackrel{v\rightarrow0}{=}&
k_4\frac{(x+y)^{-1-\e}}{(4\pi)^{2-\e}}
\left\{\left(\ln{\left(\frac{v}{1+v}\right)}+2\ln{2}\right)
\left[-2-\frac{4}{3}v+\frac{4}{15}v^2-\frac{4}{35}v^3+\frac{4}{63}v^4
+\ldots\right]\right.\nonumber\\&&\left.
+4+\frac{20}{9}v-\frac{124}{225}v^2+\frac{988}{3675}v^3
-\frac{3244}{19845}v^4+\ldots\right\},\label{eq:a4yexp}\\
A^4(x,y)&\stackrel{z\rightarrow0}{=}&
k_4\frac{(x+y)^{-1-\e}}{(4\pi)^{2-\e}}
\left\{\ln{2}\left[4+\frac{8}{3}z-\frac{8}{15}z^2+\frac{8}{35}z^3
-\frac{8}{63}z^4+\ldots\right]-\frac{2}{3}z-\frac{1}{15}z^2
+\frac{8}{105}z^3-\frac{23}{378}z^4+\ldots\right\}.
\label{eq:a4xexp}\nonumber\\
\eea
The statement that the expansion of the hypergeometric functions in powers 
of $\e$ is nontrivial is fortunately not the whole story.  It is possible 
to find the full solution as $\e\rightarrow0$ by rewriting the 
hypergeometric functions in their integral representation, expanding the 
integrand and exploring whether or not it is possible to evaluate the 
resulting integrals.  In the case of using $v$ as the variable this is not 
the case -- the expression reads:
\be
A^4(x,y)=k_4\frac{(x+y)^{-1-\e}}{(4\pi)^{2-\e}}\frac{(1+v)}{\sqrt{v}}
\left\{-4\ln{2}\arctan{(\sqrt{v})}-\int_0^v\frac{dt}{\sqrt{t}(1+t)}
\ln{\left(\frac{t}{1+t}\right)}\right\}
\ee
and the integral cannot be done directly in terms of known functions 
(or to phrase it more properly, the direct result is not known to the 
authors at present although indirectly the result can be inferred from 
the following discussion).  However, with $z$ as the variable, we get
\be
A^4(x,y)=k_4\frac{(x+y)^{-1-\e}}{(4\pi)^{2-\e}}\frac{(1+z)}{\sqrt{z}}
\left\{4\ln{2}\arctan{(\sqrt{z})}
-\int_0^z\frac{dt}{\sqrt{t}(1+t)}\ln{(1+t)}\right\}
\label{eq:a4int}
\ee
and the integral is
\bea
\int_0^z\frac{dt}{\sqrt{t}(1+t)}\ln{(1+t)}&=&\pi\ln{2}
-\imath\ln{(\sqrt{z}-\imath)}\left[\ln{2}+\ln{(1+z)}
-\ln{(1-\imath\sqrt{z})}-\ha\ln{(\sqrt{z}-\imath)}\right]\nonumber\\
&&+\imath\ln{(\sqrt{z}+\imath)}\left[\ln{2}+\ln{(1+z)}
-\ln{(1+\imath\sqrt{z})}-\ha\ln{(\sqrt{z}+\imath)}\right]\nonumber\\
&&-\imath\mbox{dilog}{\left(\ha-\frac{\imath}{2}\sqrt{z}\right)}
+\imath\mbox{dilog}{\left(\ha+\frac{\imath}{2}\sqrt{z}\right)},
\eea
in terms of the dilogarithmic function \cite{abram}.  An expansion in powers 
of $z$ yields the expansion Eq.~\eq{eq:a4xexp} which serves as a useful 
check.  
Whilst it is gratifying that the integral can eventually be written in terms 
of known functions, these functions must still be evaluated at some stage.  
For this reason, the integral form Eq.~\eq{eq:a4int} and the asymptotic 
forms Eqs.~(\ref{eq:a4yexp},\ref{eq:a4xexp}) are actually of more practical 
use.

It is possible to discuss the analytic structure of $A^4$.  Clearly, it is 
ultraviolet finite.  With the expansion Eq.~\eq{eq:a4yexp} in $v=y/x$ we see 
that as $y\rightarrow0$ there is a logarithmic singularity as seen 
previously for the integral $A$.  As for the singularities in $x$, by 
rewriting the integral
\be
\frac{1}{\sqrt{z}}\int_0^z\frac{dt}{\sqrt{t}(1+t)}\ln{(1+t)}
=\int_0^1\frac{dt}{\sqrt{t}(1+zt)}\ln{(1+zt)}
\ee
and knowing the analytic properties of $\arctan{(\sqrt{z})}/\sqrt{z}$ we 
see that the singularity occurs for $z=-1$ (i.e., for $x+y=0$) with branch 
cuts extending into the timelike region.  Again, this behavior is the same 
as for $A$.

Finally let us discuss the integral $B(x,y)$ satisfying equations 
\eq{eq:btde0} and 
\eq{eq:bsde0}.  We separate the function into two parts as previously:
\be
B(x,y)=F_B(x,y)G_B(x,y)
\ee
such that $F_B$ and $G_B$ obey the following differential equations:
\bea
2x\frac{\pd F_B(x,y)}{\pd x}&=&-3F_B(x,y),\\
2y\frac{\pd F_B(x,y)}{\pd y}&=&-(1+2\e)F_B(x,y),\\
F_B(x,y)2x\frac{\pd G_B(x,y)}{\pd x}&=&(4+8\e)(x+y)^{-1}A(x,y)
-4\frac{y^{-1-\e}}{x+y}X,\\
F_B(x,y)2y\frac{\pd G_B(x,y)}{\pd y}&=&-(4+8\e)(x+y)^{-1}A(x,y)
+4\frac{y^{-1-\e}}{x+y}X.
\eea
The solution to the first pair of differential equations is
\be
F_B(x,y)=x^{-3/2}y^{-1/2-\e}
\ee
with the possible constant vanishing as before on dimensional grounds.  
The second pair of equations now reads
\bea
x^{-3/2}y^{-1/2-\e}2x\frac{\pd G_B(x,y)}{\pd x}&=&(4+8\e)(x+y)^{-1}A(x,y)
-4\frac{y^{-1-\e}}{x+y}X,\\
x^{-3/2}y^{-1/2-\e}2y\frac{\pd G_B(x,y)}{\pd y}&=&-(4+8\e)(x+y)^{-1}A(x,y)
+4\frac{y^{-1-\e}}{x+y}X.
\eea
We are now faced with a potential problem -- for general $\e$, the function 
$A(x,y)$ is itself a combination of hypergeometric functions and we have 
little chance of solving the resulting differential equations.  However, 
we can write down $A(x,y)$ for vanishing $\e$.  Again, the derivatives with 
respect to $x$ and $y$ are distinguished by a minus sign and we can rewrite 
the two equations as a first order differential equation in $z=x/y$.  In 
this case, we cannot use the variable $v$ since this leads to non-integrable 
singularities.  After expanding in powers of $\e$, the equation is
\bea
\frac{dG_B(z)}{dz}&=&\frac{2(1+2\e)}{(4\pi)^{2-\e}}z^{1/2}(1+z)^{-2}
\left[-2\left(\frac{1}{\e}-\ga\right)-4\ln{2}+4\ln{(1+z)}+{\cal O}(\e)\right]
\nonumber\\
&&-\frac{2}{(4\pi)^{2-\e}}z^{1/2}(1+z)^{-1}\left[\frac{1}{\e}-\ga-2\ln{2}
+{\cal O}(\e)\right].
\eea
Since there is no interference between the integration over $z$ and the 
expansion in $\e$, we do not need to consider the terms of ${\cal O}(\e)$ 
further.  We thus obtain
\bea
\lefteqn{G_B(z)}\nonumber\\&&
=\frac{1}{(4\pi)^{2-\e}}\left\{-\frac{4z^{3/2}}{(1+z)}\left(\frac{1}{\e}
-\ga-2\ln{2}\right)+8(1+2\ln{2})\left[\frac{\sqrt{z}}{(1+z)}
-\arctan{(\sqrt{z})}\right]
+8\int_0^z\frac{dt\,\sqrt{t}}{(1+t)^2}\ln{(1+t)}+{\cal O}(\e)\right\}
\nonumber\\
\eea
where the constant $G_B(0)$ vanishes as before.  Using integration by parts, 
we have that
\be
\int_0^z\frac{dt\,\sqrt{t}}{(1+t)^2}\ln{(1+t)}
=\frac{1}{2}\int_0^z\frac{dt}{\sqrt{t}(1+t)}\ln{(1+t)}
-\frac{\sqrt{z}}{(1+z)}\ln{(1+z)}-\frac{\sqrt{z}}{(1+z)}+\arctan{(\sqrt{z})}
\ee
and so, the full solution can be written
\bea
B(x,y)&=&\frac{(x+y)^{-\e}y^{-2}}{(4\pi)^{2-\e}}
\left\{-4(1+z)^{-1}\left(\frac{1}{\e}-\ga-2\ln{2}+\ln{(1+z)}\right)
+16\ln{2}\left[\frac{1}{z(1+z)}-\frac{1}{z^{3/2}}\arctan{(\sqrt{z})}\right]
\right.\nonumber\\&&\left.
-\frac{8}{z(1+z)}\ln{(1+z)}+\frac{4}{z}\int_0^1\frac{dt}{\sqrt{t}(1+zt)}
\ln{(1+zt)}+{\cal O}(\e)\right\}.
\label{eq:bsol}
\eea
The factor $y^{-2}$ is extracted globally for convenience as it happens that 
the integral $B(x,y)$ is multiplied by $y^2$ when forming the gluon 
polarization.  Without the factor $y^{-2}$, the expression is finite as 
$y\rightarrow0$.  The singularities in $x$ occur for $z=-1$ as before, with 
branch cuts extending into the timelike region.  An expansion around $z=0$ 
is possible and reads
\bea
B(x,y)&\stackrel{z\rightarrow0}{=}&\frac{(x+y)^{-\e}y^{-2}}{(4\pi)^{2-\e}}
\left\{-4\left(\frac{1}{\e}-\ga\right)\left[1-z+z^2-z^3+z^4+\ldots\right]
+8\ln{2}\left[-\frac{1}{3}+\frac{3}{5}z-\frac{5}{7}z^2+\frac{7}{9}z^3
-\frac{9}{11}z^4+\ldots\right]
\right.\nonumber\\&&\left.
+\left[-\frac{16}{3}+\frac{28}{5}z-\frac{46}{7}z^2+\frac{202}{27}z^3
-\frac{91}{11}z^4+\ldots\right]\right\}.
\label{eq:bxexp}
\eea
Comparing Eq.~\eq{eq:bsol} with Eq.~\eq{eq:a4int} and using the known 
expansion for 
$A_4(x,y)$ for small $v$, Eq.~\eq{eq:a4yexp}, we can also expand $B(x,y)$ 
and 
the result reads:
\bea
B(x,y)&\stackrel{v\rightarrow0}{=}&\frac{(x+y)^{-\e}y^{-2}}{(4\pi)^{2-\e}}
\left\{-4\left(\frac{1}{\e}-\ga\right)\left[v-v^2+v^3-v^4+\ldots\right]
\right.\nonumber\\&&\left.
+4\left(\ln{\left(\frac{v}{1+v}\right)}+2\ln{2}\right)\left[v+v^2
-\frac{5}{3}v^3+\frac{7}{5}v^4+\ldots\right]-16v^2+\frac{64}{9}v^3
-\frac{368}{75}v^4+\ldots\right\}.
\label{eq:byexp}
\eea

\begin{figure}[t]
\includegraphics[width=0.39\linewidth]{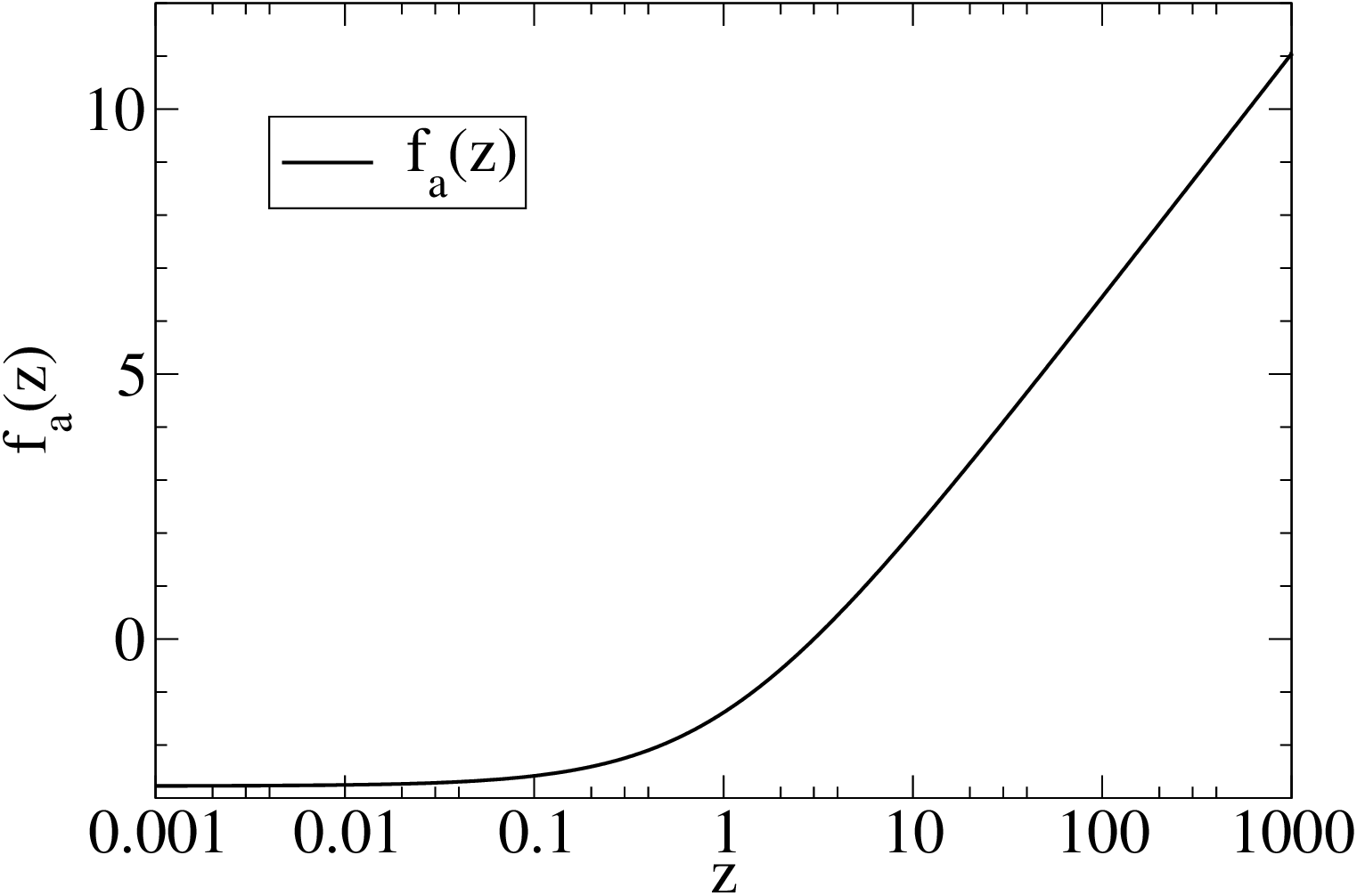}
\hspace{0.5cm}\includegraphics[width=0.37\linewidth]{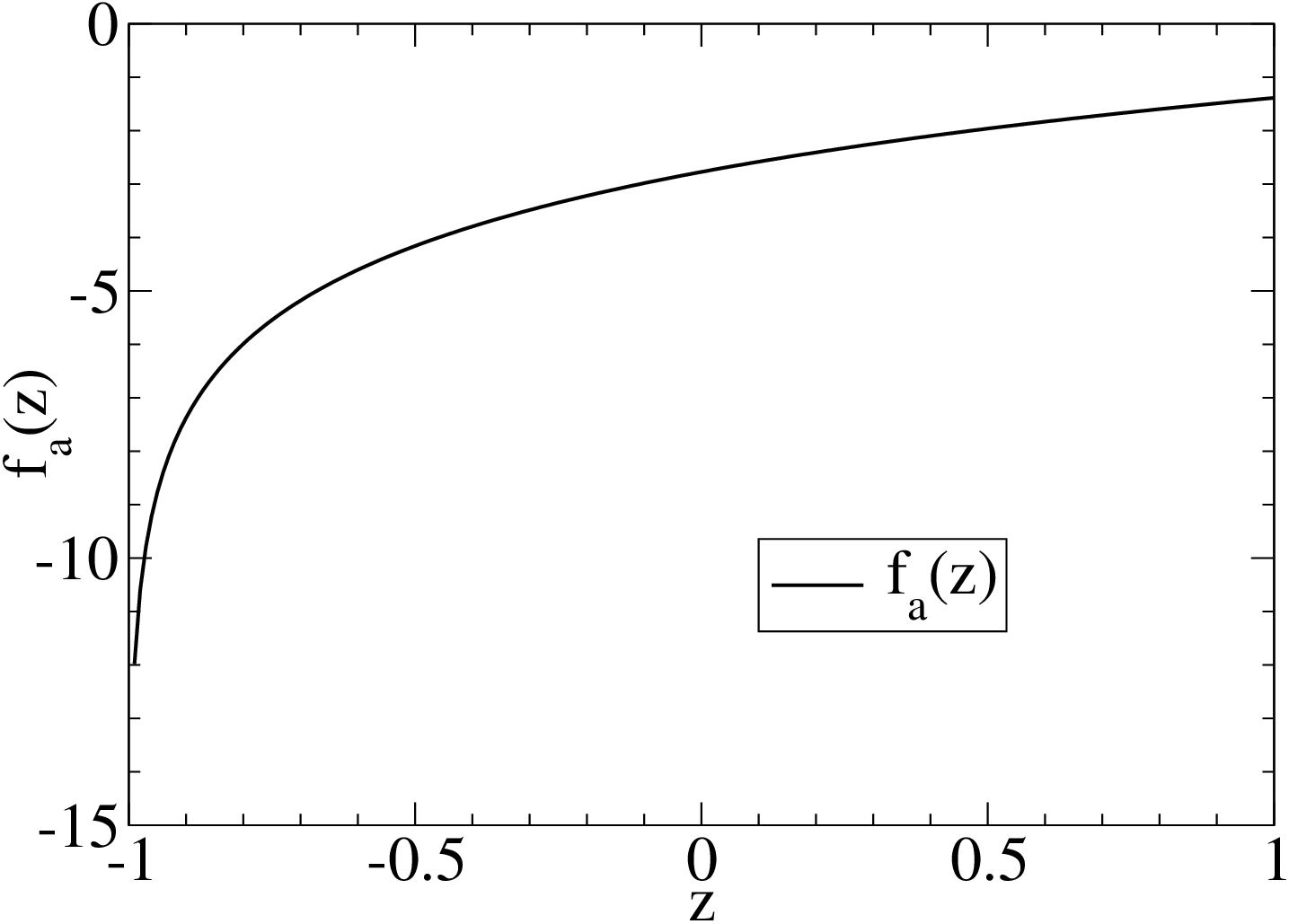}
\caption{\label{fig:afn}UV-finite part of the function $f_a(z)$.  
Left panel: Euclidean values of $z$.  Right panel: continuation of 
$f_a(z)$ into the spacelike Minkowski region.}
\end{figure}

\begin{figure}[t]
\vspace{0.5cm}
\includegraphics[width=0.39\linewidth]{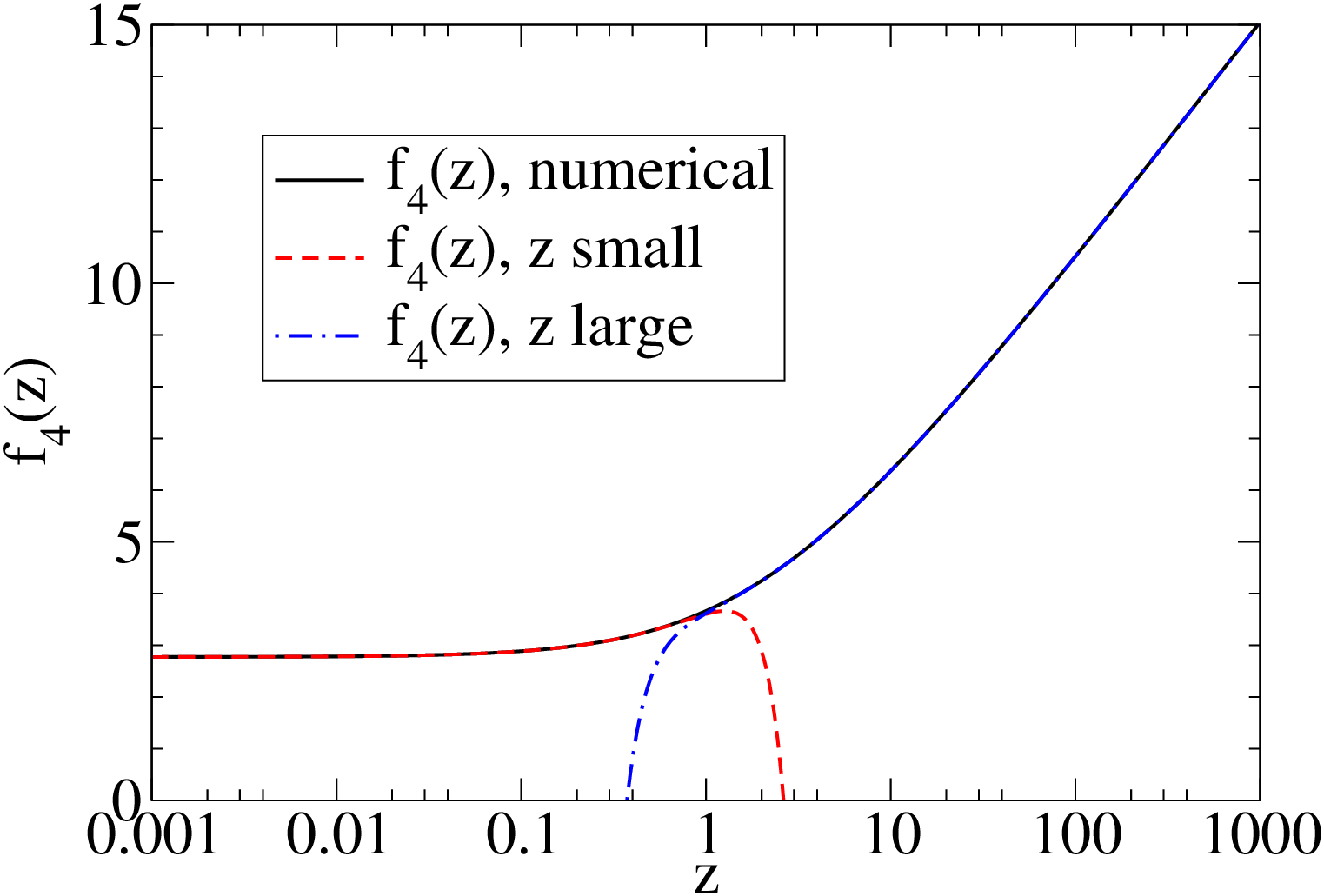}\hspace{0.5cm}
\includegraphics[width=0.37\linewidth]{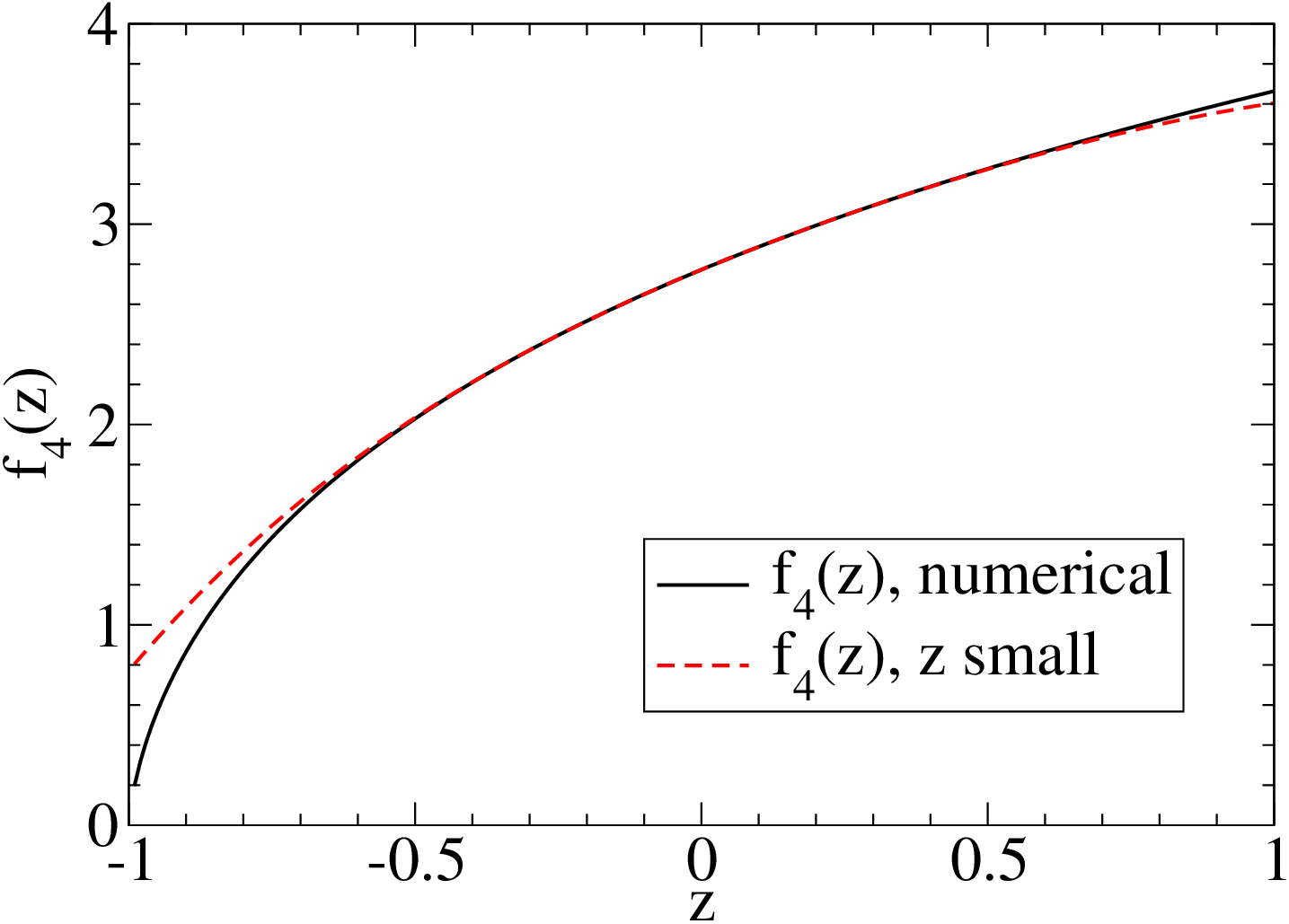}
\caption{\label{fig:4fn}The function $f_4(z)$ and its asymptotic expansions 
for $z\rightarrow0$ and $v=1/z\rightarrow0$.  Left panel: Euclidean values 
of $z$.  Right panel: continuation of $f_4(z)$ into the spacelike Minkowski 
region.}
\end{figure}

\begin{figure}[t]
\vspace{0.5cm}
\includegraphics[width=0.39\linewidth]{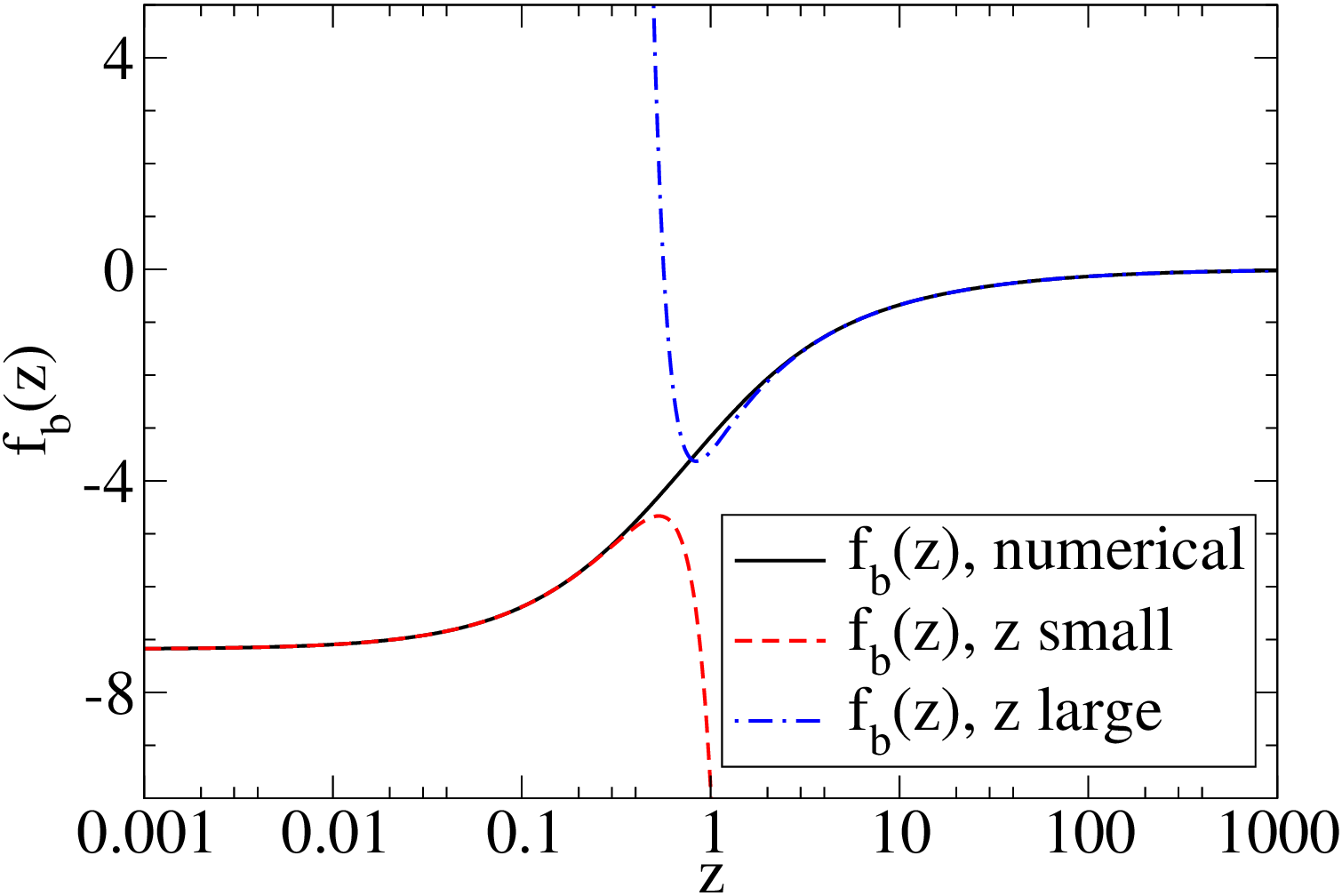}\hspace{0.5cm}
\includegraphics[width=0.37\linewidth]{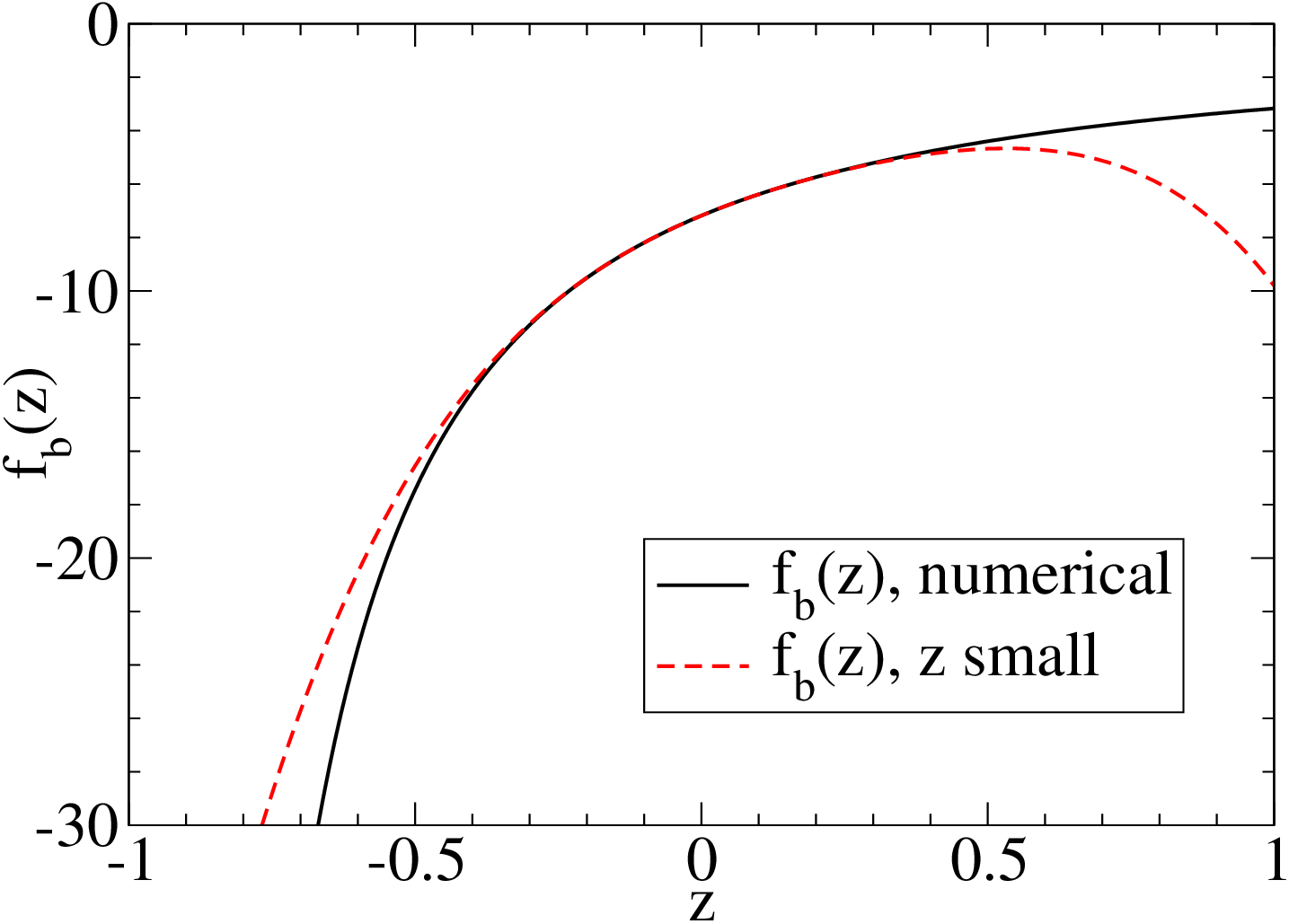}
\caption{\label{fig:bfn}UV-finite part of the function $f_b(z)$ and its 
asymptotic expansions for $z\rightarrow0$ and $v=1/z\rightarrow0$.  Left 
panel: Euclidean values of $z$.  Right panel: continuation of $f_b(z)$ into 
the spacelike Minkowski region.}
\end{figure}

Having derived the integrals, it is pertinent to see if they can be checked 
using standard techniques.  Since this is a somewhat technical exercise that 
does not add to the discussion here, we present the details in 
Appendix~\ref{app:intcheck}.  It is also useful to plot the integrals and 
their asymptotic expansions.  The functions $f_i(z)$, defined in 
Eqs.~(\ref{eq:afn}-\ref{eq:bfn}), and with the factors proportional to 
$(1/\e-\ga)$ removed are presented in Figures~\ref{fig:afn}-\ref{fig:bfn}.  
All functions are monotonically increasing with $z$.  One can see that the 
asymptotic expansions do indeed represent the functions within their domains 
of applicability.  The functions $f_a(z)$ and $f_4(z)$ are logarithmically 
singular as $y\rightarrow0$ ($z\rightarrow\infty$ with $x$ fixed), whereas 
$f_b(z)$ is finite.  All functions exhibit singularities as $z\rightarrow-1$ 
(especially strong in the case of $f_b$ although this can be attributed to 
the different choice of prefactor in the full integral $B$).  However most 
importantly, there are no singularities at $z=0$ (or equivalently $x=0$ for 
finite $y$), such that the continuation to spacelike Minkowski space is 
entirely justified.

\section{One-loop Two-Point Functions}
\setcounter{equation}{0}
Having now evaluated all the integrals that occur (see the previous section 
and Appendix~\ref{app:int0}), we can now return to the expressions for the 
one-loop, proper two-point functions and write out our results.  To do this, 
it is convenient to define two combinations of functions:
\bea
f(z)&=&4\ln{2}\frac{1}{\sqrt{z}}\arctan{\sqrt{z}}
-\int_0^1\frac{dt}{\sqrt{t}(1+zt)}\ln{(1+zt)},\nonumber\\
g(z)&=&2\ln{2}-\ln{(1+z)}.
\eea
[These functions are actually variations of the finite parts of $f_4$ and 
$f_a$ from before.]  The results are:
\bea
\G_{A\pi}^{(1)}(x,y)&=&\ov{\G}_{A\pi}^{(1)}(x,y)=0,
\label{eq:gpa0}\\
\G_{\pi\si}^{(1)}(x,y)&=&\G_c^{(1)}(y)=\frac{N_c}{(4\pi)^{2-\e}}
\left\{-\frac{4}{3}\left[\frac{1}{\e}-\ga
-\ln{\left(\frac{y}{\mu}\right)}\right]-\frac{28}{9}+\frac{8}{3}\ln{2}
+{\cal O}(\e)\right\},
\label{eq:ggh0}\\
\G_{\pi\pi}^{(1)}(x,y)&=&\frac{N_c}{(4\pi)^{2-\e}}\left\{\frac{4}{3}
\left[\frac{1}{\e}-\ga-\ln{\left(\frac{y}{\mu}\right)}\right]+\frac{52}{9}
-\frac{8}{3}\ln{2}+{\cal O}(\e)\right\},
\label{eq:gpp0}\\
\ov{\G}_{\pi\pi}^{(1)}(x,y)&=&\frac{N_c}{(4\pi)^{2-\e}}\left\{-\frac{8}{3}
+{\cal O}(\e)\right\},
\label{eq:govpp0}\\
\G_{A\si}^{(1)}(x,y)&=&\G_{\si\si}^{(1)}(x,y)=-\frac{y}{x}\ov{\G}_{AA}(x,y)
\nonumber\\
&=&\frac{N_c}{(4\pi)^{2-\e}}\left\{-\frac{1}{3}\left[\frac{1}{\e}-\ga
-\ln{\left(\frac{x+y}{\mu}\right)}\right]+\frac{1}{9}-6(1+z)+3(1+z)g(z)
+\ha(1+z)(1+3z)f(z)+{\cal O}(\e)\right\},\nonumber\\
\label{eq:gas0}\\
\G_{AA}^{(1)}(x,y)&=&\frac{N_c}{(4\pi)^{2-\e}}\left\{\left(-1
+\frac{1}{3}z\right)\left[\frac{1}{\e}-\ga-\ln{\left(\frac{x+y}{\mu}\right)}
\right]-\frac{52}{9}+\frac{116}{9}(1+z)-3z(1+z)+\frac{4}{3}g(z)
\right.\nonumber\\&&\left.
+(1+z)\left(\frac{1}{2z}-6+\frac{3}{2}z\right)g(z)+(1+z)\left(-\frac{1}{4z}
+\frac{1}{4}-\frac{11}{4}z+\frac{3}{4}z^2\right)f(z)+{\cal O}(\e)\right\}.
\label{eq:gaa0}
\eea
It is immediately apparent that for finite $y$, there are no singularities 
in any of the above at $x=0$ (or equivalently, $z=0$) since 
$f(0)=2g(0)=4\ln{2}$ cancels the $1/z$ pole.  This is to be expected since 
none of the individual integrals or their prefactors are singular at this 
point.

Let us now construct the one-loop propagator dressing functions.  Writing 
$D_{\al\ba}=D_{\al\ba}^{(0)}+g^2D_{\al\ba}^{(1)}$ and using 
Eqs.~(\ref{eq:dsepa2},\ref{eq:dseps2},\ref{eq:dsess3}), Eq.~\eq{eq:legtran} 
becomes
\bea
D_{AA}^{(1)}(x,y)&=&\G_{\pi\pi}^{(1)}(x,y)-\frac{1}{(1+z)}
\left[\G_{AA}^{(1)}(x,y)+\G_{\pi\pi}^{(1)}(x,y)\right],\nonumber\\
D_{\pi\pi}^{(1)}(x,y)&=&\G_{AA}^{(1)}(x,y)-\frac{1}{(1+z)}
\left[\G_{AA}^{(1)}(x,y)+\G_{\pi\pi}^{(1)}(x,y)\right],\nonumber\\
D_{A\pi}^{(1)}(x,y)&=&-\frac{1}{(1+z)}\left[\G_{AA}^{(1)}(x,y)
+\G_{\pi\pi}^{(1)}(x,y)\right],\nonumber\\
D_{\si\si}^{(1)}(x,y)&=&\G_{A\si}^{(1)}(x,y)-3\G_c^{(1)}(y),\nonumber\\
D_{\phi\phi}^{(1)}(x,y)&=&-\G_{A\si}^{(1)}(x,y),\nonumber\\
D_{\si\phi}^{(1)}(x,y)&=&\G_{A\si}^{(1)}(x,y)-\G_c^{(1)}(y),\nonumber\\
D_c^{(1)}(y)&=&-\G_c^{(1)}(y),\nonumber\\
D_{\si\la}^{(1)}(x,y)&=&-\G_c^{(1)}(y),\nonumber\\
D_{\phi\la}^{(1)}(x,y)&=&0.
\eea
The last equation of \eq{eq:legtran} becomes an identity by virtue of the 
relation Eq.~\eq{eq:dseovaa3}, as it should.  Putting in the above results 
for 
the one-loop, two-point proper functions, we have (in Euclidean space and 
in the limit $\e\rightarrow0$):
\bea
D_{AA}^{(1)}(x,y)&\!\!=\!\!&\frac{N_c}{(4\pi)^{2-\e}}
\left\{\left[\frac{1}{\e}-\ga-\ln{\left(\frac{x+y}{\mu}\right)}\right]
-\frac{64}{9}+3z+\left[-\frac{1}{2z}+\frac{14}{3}-\frac{3}{2}z\right]g(z)
+\left[\frac{1}{4z}-\frac{1}{4}+\frac{11}{4}z
-\frac{3}{4}z^2\right]f(z)\right\},\nonumber\\
\\
D_{\pi\pi}^{(1)}(x,y)&\!\!=\!\!&\frac{N_c}{(4\pi)^{2-\e}}
\left\{\left(-\frac{4}{3}+\frac{1}{3}z\right)\left[\frac{1}{\e}-\ga
-\ln{\left(\frac{x+y}{\mu}\right)}\right]-\frac{52}{9}+\frac{116}{9}-3z^2
+\left[\frac{11}{6}-6z+\frac{3}{2}z^2\right]g(z)
\right.\nonumber\\&&\left.
+\left[-\frac{1}{4}-\frac{1}{4}z-\frac{11}{4}z^2
+\frac{3}{4}z^3\right]f(z)\right\},\\
D_{A\pi}^{(1)}(x,y)&\!\!=\!\!&\frac{N_c}{(4\pi)^{2-\e}}
\left\{-\frac{1}{3}\left[\frac{1}{\e}-\ga
-\ln{\left(\frac{x+y}{\mu}\right)}\right]-\frac{116}{9}+3z
+\left[-\frac{1}{2z}+6-\frac{3}{2}z\right]g(z)
\right.\nonumber\\&&\left.
+\left[\frac{1}{4z}-\frac{1}{4}+\frac{11}{4}z
-\frac{3}{4}z^2\right]f(z)\right\},\\
D_{\si\si}^{(1)}(x,y)&\!\!=\!\!&\frac{N_c}{(4\pi)^{2-\e}}
\left\{\frac{11}{3}\left[\frac{1}{\e}-\ga
-\ln{\left(\frac{x+y}{\mu}\right)}\right]+\frac{31}{9}-6z+(-1+3z)g(z)
+\ha(1+z)(1+3z)f(z)\right\},\\
D_{\phi\phi}^{(1)}(x,y)&\!\!=\!\!&\frac{N_c}{(4\pi)^{2-\e}}
\left\{\frac{1}{3}\left[\frac{1}{\e}-\ga
-\ln{\left(\frac{x+y}{\mu}\right)}\right]+\frac{53}{9}+6z-3(1+z)g(z)
-\ha(1+z)(1+3z)f(z)\right\},\\
D_{\si\phi}^{(1)}(x,y)&\!\!=\!\!&\frac{N_c}{(4\pi)^{2-\e}}
\left\{\left[\frac{1}{\e}-\ga-\ln{\left(\frac{x+y}{\mu}\right)}\right]
-\frac{25}{9}-6z+\left[\frac{5}{3}+3z\right]g(z)
+\ha(1+z)(1+3z)f(z)\right\},\\
D_c^{(1)}(y)&\!\!=\!\!&D_{\si\la}^{(1)}(x,y)=\frac{N_c}{(4\pi)^{2-\e}}
\left\{\frac{4}{3}\left[\frac{1}{\e}-\ga
-\ln{\left(\frac{y}{\mu}\right)}\right]+\frac{28}{9}
-\frac{8}{3}\ln{2}\right\},\\
D_{\phi\la}^{(1)}(y)&\!\!=\!\!&0.
\eea
A few remarks are in order here.  The momentum dependence of the 
relationship between the vector propagators and proper two-point functions, 
Eq.~\eq{eq:legtran}, is such that the only occurrence of a momentum dependent 
UV-divergence is within the $D_{\pi\pi}$ propagator -- the factor $1/(1+z)$ 
is otherwise canceled within the combination 
$\G_{AA}^{(1)}+\G_{\pi\pi}^{(1)}$.  Indeed, the divergence structure of the 
one-loop, proper two-point functions has been known for some time 
\cite{Zwanziger:1998ez}.  The momentum dependent coefficient of the 
$1/\e$-pole in $D_{\pi\pi}^{(1)}$ is symptomatic of the fact that the 
$\pi$-field is not multiplicatively renormalizable since the $\pi$-field 
has its origins in the linearization of the (composite) chromoelectric 
field term of the action that is central to the first order formalism.  
However, we are further able to see that for the UV-finite parts, the 
kinematical singularities on the light-cone ($z=-1$) reside purely in the 
logarithmic term and the functions $f(z)$ and $g(z)$ which are logarithmic 
in character.  There are no singularities in the Euclidean or spacelike 
Minkowski regions ($z>-1$).  Hence, we can conclude that the analytic 
continuations between Euclidean and Minkowski space have entirely the same 
character as in linear covariant gauges -- that is to say that the 
continuation is justified.

Although it is not our intention to discuss the renormalization aspects of 
the two-point Green's functions in Coulomb gauge, at the one-loop 
perturbative level it is possible to identify two renormalization group 
invariant combinations of propagator dressing functions via the coefficients 
of the $1/\e$ poles.  In Landau gauge, a renormalization group invariant 
running coupling may be defined through the combination of gluon and ghost 
propagator dressing functions: $g^2D_{AA}D_c^2$, \cite{Fischer:2006ub}.  
This stems from the Slavnov--Taylor identity which expresses the 
universality of the coupling and the Landau gauge property that the 
ghost-gluon vertex is UV-finite.  Given that at one-loop, the coefficient 
of the $1/\e$ pole of $g^2$ is the first coefficient of the $\ba$-function 
($b_0=-11N_c/3(4\pi)^{2-\e}$) and is gauge invariant, the combination 
$D_{AA}D_c^2$ in Landau gauge has a $1/\e$ pole with the coefficient 
$11N_c/3(4\pi)^{2-\e}$.  In Coulomb gauge, the results above clearly show 
the same result.  However, the individual coefficients for $D_{AA}$ and 
$D_c$ are different from Landau gauge.  The second renormalization group 
invariant combination of propagator dressing functions is particular to 
Coulomb gauge and is $g^2D_{\si\si}$, \cite{Zwanziger:1998ez}.  The 
coefficient of the $1/\e$ pole in $D_{\si\si}$ above clearly confirms this.

\section{Summary and Outlook}
\setcounter{equation}{0}
A one-loop perturbative analysis of Coulomb gauge Yang-Mills theory within 
the first order formalism has been undertaken.  The various propagator and 
two-point proper dressing functions have been explicitly evaluated at this 
order.  In order to do this, dimensionally regularized results for the 
noncovariant two-point loop integrals inherent to Coulomb gauge have been 
derived using techniques based on differential equations and integration by 
parts identities.

The results for the two-point functions are rather interesting.  The 
dressing functions are dimensionless functions of two independent variables 
and a mass scale introduced via the regularization.  These functions can be 
split into two parts -- one unambiguously connected to the UV-divergence 
involving also the logarithmic behavior normally associated with covariant 
gauges and a second part which is a UV-finite function of the ratio of the 
temporal to spatial components of the momentum.

The analytic continuations between Minkowski and Euclidean space within the 
noncovariant setting can be justified on the grounds that the possible 
singularities occur on the light-cone, with branch cuts extending into the 
timelike Minkowski region.  It is seen explicitly that for spacelike 
Minkowski and Euclidean momenta there are no singularities in any of the 
two-point Green's functions, which is as it should be.

The outlook for future work done in Coulomb gauge is rather promising.  The 
most direct continuation of this work is to consider the vertex functions 
of the theory.  A generalization of the differential equation technique to 
the various one-loop three-point loop integrals certainly appears feasible 
albeit challenging.  Subsequently, a two-loop perturbative analysis 
studying, for example, the cancellation of potentially energy divergent 
integrals or the renormalization would be of great interest.  A second area 
of interest would be to include quarks and to study physical high energy 
processes, as has been done in linear covariant gauges.  The relationship 
between the covariant and noncovariant descriptions of the same phenomena 
will undoubtedly lead to greater insight into the physical mechanisms at 
work.

One of the motivations for studying Coulomb gauge is that nonperturbative 
phenomena such as confinement and bound states may be better understood in 
this gauge.  The analysis of nonperturbative physics is however greatly 
constrained by the perturbative behavior.  As an example, consider the 
evaluation of nonperturbative loop integrals -- the phase space of the 
integration measure still contains the perturbative domain and 
renormalization is still necessary despite the fact that one may be 
considering infrared external momentum scales.  Further, the techniques 
developed here to evaluate the perturbative integrals will almost certainly 
be of help when studying the \DS equations nonperturbatively.  Also, the 
perturbative expansion (although asymptotic) is of great use in verifying 
nonperturbative identities such as the Slavnov--Taylor identities.  The 
Slavnov--Taylor identities are the focus of present work \cite{me}.

\begin{acknowledgments}
This work has been supported by the Deutsche Forschungsgemeinschaft (DFG) 
under contracts no. DFG-Re856/6-1 and DFG-Re856/6-2.
\end{acknowledgments}

\appendix
\section{Standard Integrals}
\label{app:int0}
\setcounter{equation}{0}
There are two particular types of integral that we wish to consider in this 
appendix and both can be done using standard techniques.  We use the 
Schwinger parametrization method \cite{collins}.  The integrals considered 
have two or three denominator factors (with arbitrary powers), at least one 
of which contains both spatial and temporal components.  We list results for 
all the possible vector and tensor integrals that arise.

Consider then the integral (in Euclidean space)
\be
I=\int\frac{\dk{\w}}{\left[\w^2\right]^\mu\left[(k-\w)^2\right]^\nu}.
\ee
Using the identity \cite{abram}
\be
\frac{1}{a^\nu}=\frac{1}{\G(\nu)}\int_0^\infty d\al\,\al^{\nu-1}
\exp{\left\{-\al a\right\}},\;\;\;\;\Re{\nu}>0,\;\;\Re{a}>0
\label{eq:idgam}
\ee
we have
\be
I=\frac{1}{\G(\mu)\G(\nu)}\int_0^\infty d\al d\ba\,\al^{\mu-1}\ba^{\nu-1}
\int\dk{\w}\exp{\left\{-(\al+\ba)w^2+2\ba\s{k}{\w}-\ba k^2\right\}}.
\ee
Shifting variables
\be
\w\rightarrow\w+\frac{\ba}{\al+\ba}k
\ee
gives
\be
I=\frac{1}{\G(\mu)\G(\nu)}\int_0^\infty d\al d\ba\,\al^{\mu-1}\ba^{\nu-1}
\int\dk{\w}\exp{\left\{-(\al+\ba)w^2-\frac{\al\ba}{\al+\ba}k^2\right\}}
\ee
and rescaling, $\w\rightarrow(\al+\ba)^{-1/2}\w$, leads us to
\be
I=\frac{1}{\G(\mu)\G(\nu)}\int_0^\infty d\al d\ba\,\al^{\mu-1}\ba^{\nu-1}
(\al+\ba)^{\e-2}\exp{\left\{-\frac{\al\ba}{\al+\ba}k^2\right\}}\int\dk{\w}
\exp{\left\{-w^2\right\}}.
\ee
The integral over $\w$ can now be carried out:
\be
I=\frac{1}{(4\pi)^{2-\e}}\frac{1}{\G(\mu)\G(\nu)}\int_0^\infty d\al d\ba\,
\al^{\mu-1}\ba^{\nu-1}(\al+\ba)^{\e-2}
\exp{\left\{-\frac{\al\ba}{\al+\ba}k^2\right\}}.
\ee
By inserting the identity $1=\int_0^\infty d\la\,\de(\la-\al-\ba)$ and 
rescaling $\al\rightarrow\la\al$, $\ba\rightarrow\la\ba$ we have
\bea
I&=&\frac{1}{(4\pi)^{2-\e}}\frac{1}{\G(\mu)\G(\nu)}
\int_0^1 d\al d\ba\,\de(1-\al-\ba)\al^{\mu-1}\ba^{\nu-1}(\al+\ba)^{\e-2}
\int_0^\infty d\la\,\la^{\mu+\nu+\e-3}
\exp{\left\{-\la\frac{\al\ba}{\al+\ba}k^2\right\}},\nonumber\\
&=&\frac{1}{(4\pi)^{2-\e}}\frac{1}{\G(\mu)\G(\nu)}
\int_0^1 d\al\,\al^{\mu-1}(1-\al)^{\nu-1}
\int_0^\infty d\la\,\la^{\mu+\nu+\e-3}\exp{\left\{-\la\al(1-\al)k^2\right\}}.
\eea
The integral over $\la$ can be done and is a variation of Eq.~\eq{eq:idgam} 
giving
\be
I=\frac{1}{(4\pi)^{2-\e}}\frac{\G(\mu+\nu+\e-2)}{\G(\mu)\G(\nu)}\int_0^1 
d\al\,\al^{\mu-1}(1-\al)^{\nu-1}\left[\al(1-\al)k^2\right]^{2-\mu-\nu-\e}.
\ee
Finally the integral over $\al$ can be done (it has the integral form of 
the beta-function) to give the final result:
\be
I=\int\frac{\dk{\w}}{\left[\w^2\right]^\mu\left[(k-\w)^2\right]^\nu}
=\frac{\left[k^2\right]^{2-\mu-\nu-\e}}{(4\pi)^{2-\e}}
\frac{\G(\mu+\nu+\e-2)}{\G(\mu)\G(\nu)}
\frac{\G(2-\mu-\e)\G(2-\nu-\e)}{\G(4-\mu-\nu-2\e)}.
\ee

Now let us consider the vector integral
\be
I=\int\frac{\dk{\w}\,\w_4}{\left[\w^2\right]^\mu\left[(k-\w)^2\right]^\nu}.
\ee
In order to proceed we notice the following.  Under the change of variables 
$\w\rightarrow\w+k\ba/(\al+\ba)$ we have:
\be
\int\dk{\w}\exp{\left\{-(\al+\ba)w^2+2\ba\s{k}{\w}-\ba k^2\right\}}
=\int\dk{\w}\exp{\left\{-(\al+\ba)w^2-\frac{\al\ba}{\al+\ba}k^2\right\}}.
\label{eq:vecint}
\ee
Differentiating with respect to $k_4$ gives
\be
2\ba\int\dk{\w}(\w_4-k_4)
\exp{\left\{-(\al+\ba)w^2+2\ba\s{k}{\w}-\ba k^2\right\}}
=-2\ba\int\dk{\w}\frac{\al}{\al+\ba}k_4
\exp{\left\{-(\al+\ba)w^2-\frac{\al\ba}{\al+\ba}k^2\right\}}
\ee
which shows us that
\be
\int\dk{\w}\,\w_4\exp{\left\{-(\al+\ba)w^2+2\ba\s{k}{\w}-\ba k^2\right\}}
=\int\dk{\w}\frac{\ba}{\al+\ba}k_4
\exp{\left\{-(\al+\ba)w^2-\frac{\al\ba}{\al+\ba}k^2\right\}}.
\ee
Further differentiation gives rise to expressions for integrals involving 
other numerator structures.  Proceeding as before, we have the results
\bea
\int\frac{\dk{\w}\,\w_4}{\left[\w^2\right]^\mu\left[(k-\w)^2\right]^\nu}
&=&k_4\frac{\left[k^2\right]^{2-\mu-\nu-\e}}{(4\pi)^{2-\e}}
\frac{\G(\mu+\nu+\e-2)}{\G(\mu)\G(\nu)}
\frac{\G(3-\mu-\e)\G(2-\nu-\e)}{\G(5-\mu-\nu-2\e)},
\\
\int\frac{\dk{\w}\,\w_i}{\left[\w^2\right]^\mu\left[(k-\w)^2\right]^\nu}
&=&k_i\frac{\left[k^2\right]^{2-\mu-\nu-\e}}{(4\pi)^{2-\e}}
\frac{\G(\mu+\nu+\e-2)}{\G(\mu)\G(\nu)}
\frac{\G(3-\mu-\e)\G(2-\nu-\e)}{\G(5-\mu-\nu-2\e)},
\\
\int\frac{\dk{\w}\,\w_4^2}{\left[\w^2\right]^\mu\left[(k-\w)^2\right]^\nu}
&=&\frac{\left[k^2\right]^{3-\mu-\nu-\e}}{(4\pi)^{2-\e}}
\frac{\G(\mu+\nu+\e-3)}{\G(\mu)\G(\nu)}
\frac{\G(3-\mu-\e)\G(2-\nu-\e)}{\G(6-\mu-\nu-2\e)}\times
\nonumber\\&&
\left\{\ha(2-\nu-\e)+\frac{k_4^2}{k^2}(\mu+\nu+\e-3)(3-\mu-\e)\right\},
\\
\int\frac{\dk{\w}\,\w_i\w_j}{\left[\w^2\right]^\mu\left[(k-\w)^2\right]^\nu}
&=&\frac{\left[k^2\right]^{3-\mu-\nu-\e}}{(4\pi)^{2-\e}}
\frac{\G(\mu+\nu+\e-3)}{\G(\mu)\G(\nu)}
\frac{\G(3-\mu-\e)\G(2-\nu-\e)}{\G(6-\mu-\nu-2\e)}\times
\nonumber\\&&
\left\{\ha\de_{ij}(2-\nu-\e)+\frac{k_ik_j}{k^2}(\mu+\nu+\e-3)
(3-\mu-\e)\right\},
\\
\int\frac{\dk{\w}\,\w_4\w_i}{\left[\w^2\right]^\mu\left[(k-\w)^2\right]^\nu}
&=&k_4k_i\frac{\left[k^2\right]^{2-\mu-\nu-\e}}{(4\pi)^{2-\e}}
\frac{\G(\mu+\nu+\e-2)}{\G(\mu)\G(\nu)}
\frac{\G(4-\mu-\e)\G(2-\nu-\e)}{\G(6-\mu-\nu-2\e)},
\\
\int\frac{\dk{\w}\,\w_4\w_i\w_j}{\left[\w^2\right]^\mu
\left[(k-\w)^2\right]^\nu}
&=&k_4\frac{\left[k^2\right]^{3-\mu-\nu-\e}}{(4\pi)^{2-\e}}
\frac{\G(\mu+\nu+\e-3)}{\G(\mu)\G(\nu)}
\frac{\G(4-\mu-\e)\G(2-\nu-\e)}{\G(7-\mu-\nu-2\e)}\times
\nonumber\\&&
\left\{\ha\de_{ij}(2-\nu-\e)+\frac{k_ik_j}{k^2}(\mu+\nu+\e-3)
(4-\mu-\e)\right\}.
\eea

Using exactly the same techniques we have
\bea
\int\frac{\dk{\w}}{\left[\w^2\right]^\mu\left[(\vec{k}-\vec{w})^2\right]^\nu}
&=&\frac{\left[\vec{k}^2\right]^{2-\mu-\nu-\e}}{(4\pi)^{2-e}}
\frac{\G(\mu+\nu+\e-2)}{\G(\mu)\G(\nu)}
\frac{\G(3/2-\nu-\e)\G(2-\mu-\e)}{\G(7/2-\mu-\nu-2\e)},
\\
\int\frac{\dk{\w}\,}{\left[\w^2\right]^\mu
\left[(\vec{k}-\vec{w})^2\right]^\nu\left[\vec{\w}^2\right]^\ro}
&=&\frac{\left[\vec{k}^2\right]^{2-\mu-\nu-\ro-\e}}{(4\pi)^{2-e}}
\frac{\G(\mu+\nu+\ro+\e-2)}{\G(\mu)\G(\nu)}
\frac{\G(\mu-1/2)}{\G(\mu+\ro-1/2)}
\frac{\G(2-\mu-\ro-\e)\G(3/2-\nu-\e)}{\G(7/2-\mu-\nu-\ro-2\e)},\nonumber\\
\\
\int\frac{\dk{\w}\,\w_4^n}{\left[\w^2\right]^\mu
\left[(\vec{k}-\vec{w})^2\right]^\nu\left[\vec{\w}^2\right]^\ro}
&=&0\;\;\;\;(\mbox{$n$, odd}),
\\
\int\frac{\dk{\w}\,\w_i}{\left[\w^2\right]^\mu
\left[(\vec{k}-\vec{w})^2\right]^\nu\left[\vec{\w}^2\right]^\ro}
&=&k_i\frac{\left[\vec{k}^2\right]^{2-\mu-\nu-\ro-\e}}{(4\pi)^{2-e}}
\frac{\G(\mu+\nu+\ro+\e-2)}{\G(\mu)\G(\nu)}
\frac{\G(\mu-1/2)}{\G(\mu+\ro-1/2)}
\frac{\G(3-\mu-\ro-\e)\G(3/2-\nu-\e)}{\G(9/2-\mu-\nu-\ro-2\e)},\nonumber\\
\\
\int\frac{\dk{\w}\,\w_i\w_j}{\left[\w^2\right]^\mu
\left[(\vec{k}-\vec{w})^2\right]^\nu\left[\vec{\w}^2\right]^\ro}
&=&\frac{\left[\vec{k}^2\right]^{3-\mu-\nu-\ro-\e}}{(4\pi)^{2-e}}
\frac{\G(\mu+\nu+\ro+\e-3)}{\G(\mu)\G(\nu)}
\frac{\G(\mu-1/2)}{\G(\mu+\ro-1/2)}
\frac{\G(3-\mu-\ro-\e)\G(3/2-\nu-\e)}{\G(11/2-\mu-\nu-\ro-2\e)}\times
\nonumber\\&&
\left\{\de_{ij}\ha(3/2-\nu-\e)+\frac{k_ik_j}{\vec{k}^2}(\mu+\nu+\ro-3+\e)
(3-\mu-\ro-\e)\right\}.
\eea

\section{Checking the Nonstandard Integrals}
\label{app:intcheck}
\setcounter{equation}{0}
Since the integrals $A$, $A^4$ and $B$ must be derived using nonstandard 
techniques, it is worthwhile checking them where possible against available 
results.  It turns out that the expansions around $z=0$, 
Eqs.~(\ref{eq:asol},\ref{eq:a4xexp},\ref{eq:bxexp}), may be checked 
analytically.  An expansion around $v=0$ is not possible, since all 
integrals are divergent as $y\rightarrow0$.

To begin, consider the integral $A$, Eq.~\eq{eq:adef}.  Using Schwinger 
parameters \cite{collins}, we can rewrite the denominator factors as 
exponentials, the result being:
\be
A=\int_0^\infty\,d\al d\ba d\ga \int\dk{\w}\exp{\left\{-(\al+\ba)\w_4^2
+2\ba k_4\w_4-\ba k_4^2-(\al+\ba+\ga)\vec{\w}^2+2\ba\s{\vec{k}}{\vec{\w}}
-\ba\vec{k}^2\right\}}.
\ee
Changing variables
\be
\w_4\rightarrow\w_4+\frac{\ba}{\al+\ba}k_4,\;\;\;\;
\vec{\w}\rightarrow\vec{\w}+\frac{\ba}{\al+\ba+\ga}\vec{k}
\ee
completes the squares to give
\be
A=\int_0^\infty\,d\al d\ba d\ga \int\dk{\w}\exp{\left\{-(\al+\ba)\w_4^2
-\frac{\al\ba}{\al+\ba}x-(\al+\ba+\ga)\vec{\w}^2
-\frac{(\al+\ga)\ba}{\al+\ba+\ga}y\right\}}.
\ee
Scaling the integration variables 
$\w_4\rightarrow(\al+\ba)^{-1/2}\w_4$, 
$\vec{\w}\rightarrow(\al+\ba+\ga)^{-1/2}\vec{\w}$ then allows us to do the 
momentum integration, leaving the parametric integral
\be
A=\frac{1}{(4\pi)^{2-\e}}\int_0^\infty\,d\al d\ba d\ga(\al+\ba)^{-1/2}
(\al+\ba+\ga)^{\e-3/2}\exp{\left\{-\frac{\al\ba}{\al+\ba}x
-\frac{(\al+\ga)\ba}{\al+\ba+\ga}y\right\}}.
\ee
By inserting the identity $1=\int_0^\infty\,d\la\de(\la-\al-\ba-\ga)$ and 
rescaling all parameters by $\la$, we then get
\bea
A&=&\frac{1}{(4\pi)^{2-\e}}\int_0^1\,d\al d\ba 
d\ga\de(1-\al-\ba-\ga)(\al+\ba)^{-1/2}(\al+\ba+\ga)^{\e-3/2}
\int_0^\infty d\la\,\la^{\e}\exp{\left\{-\la\frac{\al\ba}{\al+\ba}x
-\la\frac{(\al+\ga)\ba}{\al+\ba+\ga}y\right\}}\nonumber\\
&=&\frac{(x+y)^{-1-\e}}{(4\pi)^{2-\e}}\G(1+\e)\int_0^1 d\ba
\int_0^{1-\ba}d\al\,(\al+\ba)^{-1/2}\left[\frac{\al\ba}{\al+\ba}
\frac{z}{(1+z)}+\ba(1-\ba)\frac{1}{(1+z)}\right]^{-1-\e}.
\eea
This last equation we denote as the parametric form of the integral.  For 
general values of $z$, it cannot be done because of the highly nontrivial 
denominator factor (and clearly which is why the differential technique has 
been developed).  However, knowing that the result is well-defined at $z=0$, 
we are able to make an expansion around this point and then do the resulting 
parametric integrals.  To second order in powers of $z$ we have:
\bea
\lefteqn{A\stackrel{z\rightarrow0}{=}\frac{(x+y)^{-1-\e}}{(4\pi)^{2-\e}}
\G(1+\e)\int_0^1 d\ba\int_0^{1-\ba}d\al\,(\al+\ba)^{-1/2}\times}\nonumber\\&&
\left\{\ba^{-1-\e}(1-\ba)^{-1-\e}+(1+\e)\ba^{-\e}(1-\ba)^{-2-\e}
\frac{1-\al-\ba}{\al+\ba}(z-z^2)
+\frac{(1+\e)(2+\e)}{2}\ba^{1-\e}(1-\ba)^{-3-\e}
\frac{(1-\al-\ba)^2}{(\al+\ba)^2}z^2\right.
\nonumber\\&&\left.\frac{}{}+{\cal O}(z^3)\right\}.
\eea
The parametric integrals can be done without difficulty and the result is
\be
A(x,y)\stackrel{z\rightarrow0}{=}\frac{(x+y)^{-1-\e}}{(4\pi)^{2-\e}}
\left\{-2\left(\frac{1}{\e}-\ga\right)-4\ln{2}+2z-z^2+{\cal O}(z^3)
+{\cal O}(\e)\right\}
\ee
which agrees explicitly with Eq.~\eq{eq:asol}.  Actually, with hindsight and 
given patience in expanding the integral, it appears that this particular 
integral would be possible by resumming the series expansion!

Turning to the integral $A^4$, Eq.~\eq{eq:a4def}, the parametric form of the 
integral can be written down almost immediately given the previous case.  
It reads:
\be
A^4=k_4\frac{(x+y)^{-1-\e}}{(4\pi)^{2-\e}}\G(1+\e)\int_0^1 d\ba
\int_0^{1-\ba}d\al\,\frac{\ba}{(\al+\ba)^{3/2}}\left[\frac{\al\ba}{\al+\ba}
\frac{z}{(1+z)}+\ba(1-\ba)\frac{1}{(1+z)}\right]^{-1-\e}.
\ee
Expanding again to second order in $z$ gives
\be
A^4(x,y)\stackrel{z\rightarrow0}{=}k_4\frac{(x+y)^{-1-\e}}{(4\pi)^{2-\e}}
\left\{4\ln{2}+\left(\frac{8}{3}\ln{2}-\frac{2}{3}\right)z
+\left(-\frac{8}{15}\ln{2}-\frac{1}{15}\right)z^2+{\cal O}(z^3)
+{\cal O}(\e)\right\}
\ee
which agrees explicitly with Eq.~\eq{eq:a4xexp}.  Although at this order it 
may 
appear reasonable to suppose that one may resum the series to recover the 
full function, a quick glance at the full expansion Eq.~\eq{eq:a4xexp} and 
the 
solution Eq.~\eq{eq:a4int} tells us otherwise.

As might be expected, the integral $B$, Eq.~\eq{eq:bdef}, is rather more 
complicated.  The parametric form reads:
\be
B=\frac{(x+y)^{-\e}y^{-2}}{(4\pi)^{2-\e}}\frac{\G(2+\e)}{(1+z)^2}
\int_0^1 d\al\int_0^{1-\al}d\ga\int_0^{1-\al-\ga}d\ba(\al+\ba)^{-1/2}
\left[\frac{\al\ba}{\al+\ba}\frac{z}{1+z}+(\al+\ga)(1-\al-\ga)
\frac{1}{1+z}\right]^{-2-\e}.
\ee
The last factor can be expanded in powers of $z$ and to second order, we get
\bea
B=\frac{(x+y)^{-\e}y^{-2}}{(4\pi)^{2-\e}}\frac{\G(2+\e)}{(1+z)^2}&\!\!\!\!
&\int_0^1 d\al\int_0^{1-\al}d\ga\int_0^{1-\al-\ga}d\ba(\al+\ba)^{-1/2}
\times\nonumber\\
&&\left\{(\al+\ga)^{-2-\e}(1-\al-\ga)^{-2-\e}\left[1+(2+\e)z
+\ha(1+\e)(2+\e)z^2+\ldots\right]
\right.\nonumber\\&&
+\frac{\al\ba}{\al+\ba}(\al+\ga)^{-3-\e}(1-\al-\ga)^{-3-\e}
\left[-(2+\e)z-(2+\e)^2z^2+\ldots\right]
\nonumber\\&&\left.
+\left(\frac{\al\ba}{\al+\ba}\right)^2(\al+\ga)^{-4-\e}(1-\al-\ga)^{-4-\e}
\left[\ha(2+\e)(3+\e)z^2+\ldots\right]\right\}.
\eea
The integral over $\ba$ can be done and gives
\bea
B&=&\frac{(x+y)^{-\e}y^{-2}}{(4\pi)^{2-\e}}\frac{\G(2+\e)}{(1+z)^2}
\int_0^1 d\al\int_0^{1-\al}d\ga\times\nonumber\\
&&\left\{2\left[(1-\ga)^{1/2}-\al^{1/2}\right]
(\al+\ga)^{-2-\e}(1-\al-\ga)^{-2-\e}\left[1+(2+\e)z+\ha(1+\e)(2+\e)z^2
+\ldots\right]
\right.\nonumber\\&&
+\left[-4\al^{3/2}+2\al^2(1-\ga)^{-1/2}+2\al(1-\ga)^{1/2}\right]
(\al+\ga)^{-3-\e}(1-\al-\ga)^{-3-\e}\left[-(2+\e)z-(2+\e)^2z^2+\ldots\right]
\nonumber\\&&
+\left[2\al^2(1-\ga)^{1/2}+4\al^3(1-\ga)^{-1/2}
-\frac{2}{3}\al^4(1-\ga)^{-3/2}-\frac{16}{3}\al^{5/2}\right]\times
\nonumber\\&&\left.
(\al+\ga)^{-4-\e}(1-\al-\ga)^{-4-\e}\left[\ha(2+\e)(3+\e)z^2
+\ldots\right]\right\}.
\eea
Now, in order to do the last pair of parametric integrals, we change 
variables with $\al=r(1-s)$, $\ga=rs$ such that now
\bea
B&=&\frac{(x+y)^{-\e}y^{-2}}{(4\pi)^{2-\e}}\frac{\G(2+\e)}{(1+z)^2}
\int_0^1 dr\,r\int_0^1ds\times\nonumber\\
&&\left\{2\left[(1-rs)^{1/2}-r^{1/2}(1-s)^{1/2}\right]r^{-2-\e}
(1-r)^{-2-\e}\left[1+(2+\e)z+\ha(1+\e)(2+\e)z^2+\ldots\right]
\right.\nonumber\\&&
+\left[-4r^{3/2}(1-s)^{3/2}+2r^2(1-s)^2(1-rs)^{-1/2}
+2r(1-s)(1-rs)^{1/2}\right]\times
\nonumber\\&&
r^{-3-\e}(1-r)^{-3-\e}\left[-(2+\e)z-(2+\e)^2z^2+\ldots\right]
\nonumber\\&&
+\left[2r^2(1-s)^2(1-rs)^{1/2}+4r^3(1-s)^3(1-rs)^{-1/2}
-\frac{2}{3}r^4(1-s)^4(1-rs)^{-3/2}
-\frac{16}{3}r^{5/2}(1-s)^{5/2}\right]\times
\nonumber\\&&\left.
r^{-4-\e}(1-r)^{-4-\e}\left[\ha(2+\e)(3+\e)z^2+\ldots\right]\right\}.
\eea
The two-dimensional integral in $\al$ and $\ga$ is now separated into two 
parts which can be done in turn.  The integral over $s$ yields powers of 
$r$ and $(1-r)$ and the integral over $r$ subsequently leads to the familiar 
combinations of gamma functions.  Expanding in $\e$ and completing the 
expansion of $z$ by including the prefactor $1/(1+z)^2$ we finally arrive at
\bea
B=\frac{(x+y)^{-\e}y^{-2}}{(4\pi)^{2-\e}}\left\{-4\left(\frac{1}{\e}
-\ga\right)\left[1-z+z^2\right]+8\ln{2}\left[-\frac{1}{3}+\frac{3}{5}z
-\frac{5}{7}z^2\right]-\frac{16}{3}+\frac{28}{5}z-\frac{46}{7}z^2
+{\cal O}(z^3)+{\cal O}(\e)\right\}\nonumber\\
\eea
which agrees explicitly with Eq.~\eq{eq:bxexp}.


\end{document}